\documentclass[a4paper, 12pt]{article}
\usepackage[utf8]{inputenc}
\usepackage[T1]{fontenc}
\usepackage{amsmath,bbm}
\usepackage[numbers,sort&compress]{natbib}
\setcitestyle{citesep={,}}
\usepackage{amsfonts}
\usepackage{mdframed}
\usepackage{amssymb,comment}
\usepackage{graphicx}
\usepackage{transparent}
\usepackage{notoccite}
\usepackage[normalem]{ulem}
\usepackage[left=2cm, right=2cm]{geometry}
\usepackage{todonotes}
\usepackage{xcolor}
\usepackage{physics}
\usepackage{soul}
\usepackage{bbold}
\usepackage{ntheorem}
\usepackage{setspace} 
\usepackage[overlay,absolute]{textpos}
\usepackage{hyperref}

\newcommand{\be}{\begin{equation}}
	\newcommand{\ee}{\end{equation}}
\newcommand{\bea}{\begin{eqnarray}}
	\newcommand{\eea}{\end{eqnarray}}

\newcommand{\nocitebrackets}[1]{\@cite{\@citex[][#1]}}
\newcommand{\plaincite}[1]{\begingroup
  \let\@cite\@firstofone
  \cite{#1}%
  \endgroup}
\definecolor{DS}{HTML}{629202}
\definecolor{AH}{HTML}{d91f05}
\definecolor{BF}{HTML}{f903d7}

\newcommand\PlaceText[3]{%
\begin{textblock*}{10in}(#1,#2)
#3
\end{textblock*}
}%
\begin{document}

\sloppy

\PlaceText{165mm}{15mm}{May 12, 2025}
\begin{center}
{\LARGE\bf Localized Gravity, de Sitter,\\[.3cm]
and the Horizon Criterion}

\vspace{.5cm}
\end{center} 

\vspace*{0.8cm}
\thispagestyle{empty}

\centerline{\large 
Bjoern Hassfeld, Arthur Hebecker, Daniel Schiller}
\vspace{0.5cm}
 
\begin{center}
{\it Institute for Theoretical Physics, Heidelberg University,}
\\
{\it Philosophenweg 19, 69120 Heidelberg, Germany}\\[0.5ex]  
\end{center} 
\vspace{-.4cm}

\vspace{.25cm}
\centerline{\small\textit{E-Mail:} 
\href{mailto:schiller@thphys.uni-heidelberg.de}{schiller@thphys.uni-heidelberg.de},
\href{mailto:hassfeld@wisc.edu}{hassfeld@wisc.edu},}
\centerline{\href{mailto:a.hebecker@thphys.uni-heidelberg.de}{a.hebecker@thphys.uni-heidelberg.de}}

\vspace*{.5cm}
\begin{abstract}
\normalsize
\vspace*{.4cm}
\noindent
Realizing de Sitter-like solutions in string theory remains challenging, prompting speculation about which specific feature might be responsible for their  inconsistency in quantum gravity.
In this work, we focus on the `Horizon Criterion', which identifies spacetimes as problematic if they exhibit cosmological horizons. 
In particular, we study the implications for spacetimes with dynamical boundaries. 
We argue that requiring inertial observers localized on an end-of-the-world (ETW) brane to be in causal contact with every other observer is too restrictive as there exist string-theoretic solutions without this property. 
Hence, if one does not want to abandon the idea of cosmological horizons being the fundamental issue with de Sitter, a refined condition is needed.
The requirement that inertial, boundary-localized observers should be in causal contact with all other observers on the same ETW brane is such an appropriate refinement.  We explore the consequences of this criterion for ETW branes whose energy density is governed by a scalar field, considering two cases: First, with a scalar field confined to the ETW brane, and second, with a bulk modulus subject to a brane-localized potential.
\end{abstract}
\vspace{10pt}

\newpage

\tableofcontents 


\section{Introduction}
It has been known for a long time that dS solutions are hard to realize in string theory. More recently, this has been turned into a Swampland Conjecture against stringy dS models \cite{Danielsson:2018ztv, Obied:2018sgi}. Renewed scrutiny of the leading stringy dS models \cite{Kachru:2003aw, Balasubramanian:2005zx} has then led to the preliminary conclusion that serious control issues exist~\cite{Carta:2019rhx, Gao:2020xqh, Junghans:2022exo, Gao:2022fdi, Junghans:2022kxg, Hebecker:2022zme, Schreyer:2022len, Schreyer:2024pml}. See also \cite{Demirtas:2021nlu, Hebecker:2020ejb, Krippendorf:2023idy, McAllister:2024lnt, Lanza:2024uis, Chauhan:2025rdj} for recent work aimed at improving these leading dS constructions. Independently of such stringy issues, arguments against the consistency of dS space have been discussed for decades, see e.g.~\cite{Ford:1984hs,Mottola:1984ar,Antoniadis:1985pj,Tsamis:1992sx,Goheer:2002vf, Polyakov:2007mm, Dvali:2014gua, Dvali:2017eba}.

Let us for the moment assume that string theory or even quantum gravity in general do indeed abhor dS space. If so, the question is which particular feature of this cosmological solution makes it problematic. A fairly obvious possibility is the presence of cosmological horizons \cite{Fischler:2001yj, Hellerman:2001yi,Bedroya:2019snp}.
The option of forbidding them has recently been revived in the Swampland context \cite{Hebecker:2023qke,Andriot:2023wvg}. We will call the corresponding condition the `Horizon Criterion', to be specified more precisely momentarily.
The point of the present paper is to focus on this condition and to investigate what it implies for the allowed geometries and scalar potentials. Of course, this is closely connected with the swampland-type question about accelerated expansion in asymptotic regimes of string compactifications \cite{Ooguri:2018wrx, Hebecker:2018vxz,Andriot:2018mav} and, more specifically, about precise bounds on the underlying asymptotic potentials \cite{Rudelius:2021oaz, Rudelius:2021azq, Rudelius:2022gbz, Andriot:2022xjh,
Calderon-Infante:2022nxb, Marconnet:2022fmx, Shiu:2023nph, Shiu:2023fhb,Cremonini:2023suw,Freigang:2023ogu,VanRiet:2023cca,Marconnet:2025vhj} (see also \cite{Garg:2018reu}). For earlier studies of eternal cosmic acceleration in the string theory context see e.g.~\cite{Townsend:2003fx,Ohta:2003pu,Chen:2003dca,Andersson:2006du,Tsujikawa:2013fta,Russo:2018akp,Grimm:2019ixq}.
However, as we discuss at the end of Sect.~\ref{Sect:Localized_Universes}, the corresponding `no asymptotic acceleration conjecture' is hard to make suitably precise or phrase as a fundamental quantum gravity constraint. In our subsequent study of the `Horizon Criterion' we will be particularly interested in spacetimes with boundaries, including but not limited to models with localized gravity \cite{Randall:1999ee, Randall:1999vf, Kaloper:1999sm, Karch:2000ct}, as recently discussed in the Swampland context in \cite{Anastasi:2025puv}.\footnote{The idea of dynamical spacetime boundaries or `End-of-the-World' branes first appeared in \cite{Jourjine:1983du} and are expected to be ubiquitous in quantum gravity thanks to the Cobordism conjectures \cite{McNamara:2019rup}. For a selection of recent related studies see \cite{Buratti:2021yia, Blumenhagen:2022mqw, Friedrich:2023tid,Sugimoto:2023oul, Angius:2024pqk, Muntz:2024joq,Barbon:2025vvh}.}

To get started, let us make precise what we believe to be a widely accepted definition of horizons in the present context:


\vspace{0.2cm}
\begin{mdframed}[backgroundcolor=white,shadow=true,shadowsize=4pt,shadowcolor=black,roundcorner=6pt]
\textbf{Cosmological horizons:} Consider a spacetime $\mathcal{M}$ and an inertial observer with worldline $\sigma$ reaching infinite proper time. If there exist points $p\in \mathcal{M}$ such that $\sigma$ does not intersect the domain of future influence\footnote{
We refrain from writing `forward light cone' since we do not want black hole horizons to fall under our definition. In the black hole case, we assume that due to unitary evaporation the region of future influence of points inside the horizon is not limited by the singularity.
} 
of $p$ we will say that `$p$ is screened from the observer by a cosmological horizon.' This horizon is hence the boundary between points which are screened and those which are not.
Here, an observer is considered inertial if its worldline is a geodesic. In case the observer is localized to the spacetime boundary, the worldline must be a geodesic with respect to the metric induced on the boundary.
\end{mdframed}
\vspace{0.2cm}

\noindent
Based on this definition, one may formulate the `Horizon Criterion' by claiming that string theory does not allow for solutions with cosmological horizons. To illustrate this, consider $d$-dimensional FLRW spacetimes,
\begin{equation}
    ds^2_d = -d\tau^2 + a(\tau)^2 ds_{d-1}^2\,,\label{intro_FLRW_metric}
\end{equation}
where $ds_{d-1}^2$ is the maximally symmetric spatial part of the metric. The condition for the existence of a horizon then reads
\begin{align}
    \int_{\tau_0}^\infty \frac{d\tau}{a(\tau)}<\infty\,.\label{horizon_calc}
\end{align}
This implies that light-rays traverse asymptotically a finite coordinate range with respect to $ds_{d-1}^2$, establishing the existence of a cosmological horizon.
A necessary condition for horizon formation is that spacetime undergoes eternal accelerated expansion, characterized by $\ddot{a}(\tau) > 0$ for $\tau > \tau_0$\,.\,\footnote{
Nevertheless, as will be argued in Sect.~\ref{Sect:Localized_Universes}, we expect that horizons and not eternal acceleration represent the fundamentally problematic feature of de Sitter-like solutions.}

If the energy density of an FLRW universe is dominated by a scalar field $\phi_{\rm FLRW}$ with potential $V(\phi_{\rm FLRW})\sim \exp({-\gamma\phi_{\rm FLRW}})$, then 
asymptotic cosmic acceleration requires \cite{Rudelius:2021azq}
\begin{align}
    \gamma < \gamma_{\rm crit}\equiv \frac{2}{\sqrt{d-2}}\,,\label{horizon_criterion_bulk}
\end{align}
which in this case also leads to the presence of a horizon (as emphasized in the present context in \cite{Hebecker:2023qke, Andriot:2023wvg}). Thus, flat potentials are ruled out. As we will see, generalizing this logic to spacetimes with boundaries is non-trivial and, since boundaries are in general part of string theory solutions, more care is required in formulating a `Refined Horizon Criterion'.

\paragraph{Cosmological horizons, ETW branes, and the swampland:}
The study of swampland conjectures in the presence of boundaries is interesting both as a matter of principle and because boundaries are expected to be ubiquitous in string theory \cite{McNamara:2019rup}. If the bulk is AdS and the ETW-brane tension is appropriately adjusted, then gravity localizes at this so-called `UV brane' \cite{Randall:1999vf} and one is effectively dealing with a lower-dimensional Minkowski-space theory. This continues to hold for `de-tuned' brane-tension, with the lower-dimensional effective theory being correspondingly AdS or Minkowski \cite{Kaloper:1999sm, Karch:2000ct}. One may view this as a cutoff-version \cite{Verlinde:1999fy, Gubser:1999vj, Giddings:2000mu, Duff:2000mt, Arkani-Hamed:2000ijo} of AdS/CFT \cite{Maldacena:1997re}: The $d$-dimensional AdS bounded by a UV brane corresponds to a $(d\!-\!1)$-dimensional gravitational theory coupled to a CFT with UV-cutoff. The latter represents the bulk degrees of freedom.

Following this logic, one concludes that the swampland constraints on such spacetimes are just the ordinary swampland constraints of a $(d\!-\!1)$-dimensional gravity + CFT system. Nevertheless, it is of course interesting to check this explicitly, as proposed recently in \cite{Anastasi:2025puv}. Specifically in our context, one may consider a model with ETW-brane-localized scalar field $\phi$. Its potential is then subject to the constraints of the dS-conjecture or, if we assume horizons to be the problematic feature of dS spaces, to the Horizon Criterion.

\begin{figure}
	\centering
	\includegraphics[width=0.4\linewidth]{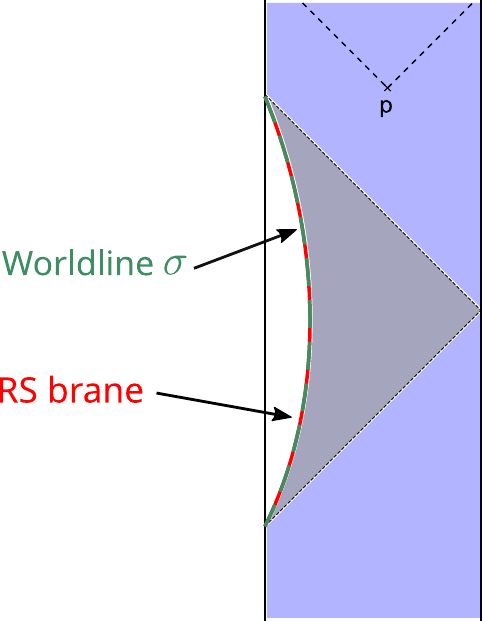}
	\caption{The Penrose diagram of $d$-dimensional AdS with a flat Randall-Sundrum UV brane embedded (red-green dashed line). The gray region corresponds to a portion of a Poincaré patch of AdS. The worldline $\sigma$ of a brane-bound observer is illustrated in green. It coincides with the location of the RS brane in this illustration, which is why the spacetime boundary is illustrated as a red-green dashed line. One observes that there exists a point p whose future domain of influence does not intersect $\sigma$.}
	\label{Fig:RS_Brane}
\end{figure}

An interesting observation can be made by considering a flat ETW brane, i.e.~a UV-brane with tension tuned as in the original Randall-Sundrum (RS) model \cite{Randall:1999vf}. We note that this model is not in the 
swampland since an explicit string-theoretic realization exists \cite{Verlinde:1999fy}. The Penrose diagram of this model is shown in Fig.~\ref{Fig:RS_Brane}. More precisely, the RS model directly corresponds to the Poincar\'e patch of AdS, colored gray in the figure, together with the brane shown in red. But this spacetime is only part of global AdS.
The UV-brane of the RS model intersects the asymptotic boundary at infinite Poincar\'e time, which is however at finite global time.  The region outside the Poincar\'e patch is inaccessible for an observer living on the brane. Thus, according to the above definition there exists a horizon.
In other words, the RS model realizes a spacetime with cosmological horizon while not being in the swampland. A swampland conjecture excluding horizons can hence not be upheld.\footnote{
A horizon also exists in the UV completion. 
To see this, recall that the Verlinde model \cite{Verlinde:1999fy} may be thought of as a compactification on $T^6/\mathbb{Z}_2$ with a stack of 16 D3 branes sitting at one point of the compact space. Replacing this stack by a warped throat, $\mathbb{R}_+\times S^5$, one discovers that the near-brane geometry is that of an $AdS_5\times S^5$ compactification from 10d to 5d. This $AdS_5\times S^5$ region may be viewed as the bulk of an RS model. The rest of the $T^6/\mathbb{Z}_2$ plays the role of the UV brane. Now, observers living in this compact space (e.g.~bound to one of the O3-planes) see a horizon precisely in the sense of Fig.~\ref{Fig:RS_Brane}. Namely, signals from the part of the $AdS_5\times S^5$ geometry above the gray region can not reach them.
}
If one wants to maintain that horizons are the underlying reason for why de Sitter space is problematic in string theory, a refined version of the Horizon Criterion is required. 

\paragraph{The Refined Horizon Criterion:}
For universes with boundary, one can distinguish between horizons for observers living in the bulk or on the brane.  
\vspace{0.2cm}
\begin{mdframed}[backgroundcolor=white,shadow=true,shadowsize=4pt,shadowcolor=black,roundcorner=6pt]
{\textbf{Brane and bulk horizons:} For a bulk observer with cosmological horizon, we will call the intersection of this horizon with the bulk a \textit{bulk-bulk} horizon. Its intersection with the ETW brane will be called its \textit{bulk-brane} horizon.  Similarly, for an observer on the ETW brane who sees a horizon, we will call the intersection of the horizon with the bulk the \textit{brane-bulk} horizon and its intersection with the ETW brane the \textit{brane-brane} horizon.}
\end{mdframed}

According to these definitions, the RS spacetime only has a brane-bulk horizon whereas the dS branes of \cite{Kaloper:1999sm,Karch:2000ct} realize a spacetime with brane-bulk {\it and} brane-brane horizons. 
{We can now formulate a refined version of the Horizon Criterion:

\vspace{0.2cm}
\begin{mdframed}[backgroundcolor=white,shadow=true,shadowsize=4pt,shadowcolor=black,roundcorner=6pt]
\textbf{Refined Horizon Criterion:} Consistent theories of quantum gravity do not allow for spacetime geometries exhibiting bulk-bulk or brane-brane horizons.
\end{mdframed}
\vspace{0.2cm}
}

\noindent
The condition for the scalar potential of brane-localized scalar that arises from this Refined Horizon Criterion is already known for ETW branes bounding AdS spacetimes: One simply makes use of AdS/CFT with cutoff and applies \eqref{horizon_criterion_bulk}. We will hence focus on studying the situation when ETW branes bound flat spacetimes.
Although gravity does not localize on such branes, they can still provide an expanding lower-dimensional geometry potentially possessing brane-brane horizons.\footnote{One might naively think that decays to nothing realize an ETW boundary with horizon. However, such decays can only occur for non-SUSY spacetimes. As one expects all Minkowski spaces in the string landscape to be supersymmetric, the only interesting case is the decay of AdS. But unstable AdS decays completely within one AdS-time, such that every observer dies in finite time and the concept of a horizon does not make sense.}
We focus on ETW branes whose energy density is set by a brane-localized scalar $\phi$ and analyze conditions on the potential $V(\phi)$ that guarantee or exclude brane-brane horizons. The dynamics of ETW branes of this type bounding AdS space has recently been considered in \cite{Fujiki:2025yyf} in the context of AdS/CFT correspondence.

In addition, we consider models where $\phi$ is a massless scalar field in the bulk, subject to a potential $V(\phi)$ localized on the ETW brane. As a motivation, one may think of $\phi$ as the string theoretic dilaton and of the ETW brane as an orientifold plane, possessing a dilaton-dependent potential. Similarly, $\phi$ may be a modulus of supersymmetric bulk, subject to a potential localized on the non-supersymmetric boundary.

For both situations, the conditions on the potential can be formulated in analogy to \eqref{horizon_criterion_bulk}.
We find that potentials of the form
\begin{align}
    V(\phi)=\begin{cases}
        \frac{c^2}{\phi^2}\,,\qquad \text{with}\qquad c<c_{\rm crit} = \frac{1}{\sqrt{2(4d-9)}} & \phi \text{ living on the brane}\\
        V_0e^{-\lambda\phi}\,,\qquad \text{with}\qquad\lambda < \lambda_{\rm crit}=\frac{1}{\sqrt{d-2}} & \phi \text{ living in the bulk}
    \end{cases} \label{intro_boundary_potentials}
\end{align}
guarantee the existence of brane-brane horizons.
In deriving these results, it is crucial to take into account that there may exist two observers on an ETW brane that are in causal contact with each other only via interactions through the spacetime bulk. 
As a result, the condition for brane-brane horizons to exist is in general not just given by \eqref{horizon_calc} with $a$ replaced by the scale factor of the induced geometry.

Furthermore, employing an autonomous system analysis (see \cite{Bahamonde:2017ize} for a review on autonomous systems in cosmology), we study the `critical cases', corresponding to potentials that to leading order take the form \eqref{intro_boundary_potentials} with $c=c_{\rm crit}\,,\lambda=\lambda_{\rm crit}$ and derive conditions on corrections to the potential that guarantee the existence of brane-brane horizons.
We also study the corresponding result for ordinary FLRW universes, refining the result \eqref{horizon_criterion_bulk}.

This paper is structured as follows. 
In Sect.~\ref{Sect:Localized_Universes}, we discuss the Refined Horizon Criterion in more detail.
In Sect.~\ref{Sect:Brane_field}, we consider a scalar field $\phi$ only living on the ETW brane and derive a general condition for brane-brane horizons to exist, in analogy to \eqref{horizon_calc}. We subsequently study which scalar field potentials realize them.
In Sect.~\ref{Sect:Bulk_field}, we consider $\phi$ to be a modulus in the bulk. 
As before, we derive a condition analogous to \eqref{horizon_calc} and study its implications on the scalar field potential on the boundary. 
In App.~\ref{App:Critical_cases}, we study solutions that marginally do (not) realize horizons in more detail. 

\section{Preliminary Reflections on Horizon Criteria}\label{Sect:Localized_Universes}
In this section, we critically reflect on Horizon Criteria as discussed in the Introduction. 
\subsection{On different types of observers}

In this subsection, we discuss the notion of an observer in more detail: We show that bulk observers in the UV theory may sometimes become brane observers from an IR perspective and critically discuss the special role of Rindler observers.

\paragraph{Brane Observer vs.~Bulk Observer:}
One may be concerned that inertial observers on an ETW brane are very different from inertial bulk observers and that a horizon criterion should only apply to the latter. As a defense against such criticism, we note that there exist situations in which the UV resolution of an ETW brane is a boundary-free higher-dimensional spacetime, such that the inertial ETW brane observer is in fact identical to an inertial bulk observer. The most prominent example of this is Witten's bubble of nothing \cite{Witten:1981gj}.
It describes the decay of the Kaluza-Klein vacuum $\mathbb{M}^{1,3}\times S^1$, i.e.~of 4d Minkowski space times an $S^1$, to nothing.\footnote{See  
\cite{GarciaEtxebarria:2020xsr}
for a string-theoretic construction of a bubble of nothing, establishing their existence in UV complete quantum gravity, and \cite{Friedrich:2023tid} for a proposal for how bubbles of nothing could naturally be part of the 4d type-IIB flux landscape.}
In the Lorentzian geometry, the $S^1$ shrinks to vanishing size over an $S^2\subset \mathbb{R}^3\subset \mathbb{R}^{1,3}$. This $S^2$ and hence the ball of `nothing' inside it grow as a function of time.
From the 4d perspective, one has an expanding ETW brane with negative tension, located on this $S^2$ \cite{Friedrich:2023tid}.
Interestingly, the worldline of an inertial observer on the ETW brane, i.e.~on the (5d smooth) locus where the $S^1$ reaches zero size, is also a geodesic in the 5d metric. Such an observer is then both an inertial ETW-brane and an inertial bulk observer. In such situations, the standard Horizon Criterion for 5d observers implies the Refined Horizon Criterion as stated in the Introduction.

\paragraph{Brane Observer vs.~Rindler Observer:} As a further concern, one may note that Rindler observers exist even in Minkowski space. They have a horizon and, as result, forbidding geometries in which certain observers have horizons appears unmotivated. Indeed, consider for example a detuned RS model in which the UV brane has a tension above the critical value. This brane accelerates outward, inducing both brane-brane and brane-bulk horizons. Similarly, a Rindler observer accelerates and induces a brane-bulk horizon. However, there is a key difference: The acceleration of the brane-bound observer originates in the ETW brane tension. 
By contrast, the acceleration of a Rindler observer requires an energy source, which the observer must carry with them and which will eventually be exhausted. As a result, eternal acceleration of the Rindler observer seems impossible -- even in principle.
We hence do not view the existence of (idealized) Rindler observers as a strong argument against our proposal.

\subsection{Potential counterexamples}
In this subsection, we address potential counterexamples to the proposed Refined Horizon Criterion.

One might naively think that horizons can be trivially obtained by considering the decay of an AdS spacetime to lower-energy AdS or to nothing. Such a proposal \cite{Banerjee:2018qey} has been developed to explain the accelerated expansion we currently observe in our universe. The domain wall or ETW brane appearing in this process may then induce brane-brane horizons. 
However, such models assume unstable AdS, and it is well-known that the latter decays completely in finite time (of order the AdS-time) \cite{Horowitz:2007pr}. 
In particular, infinitely many bubbles form in the infinite-volume region near the AdS-boundary.
Their collisions prevent the worldline of a brane-localized observer to reach the AdS-boundary, such that no Randall-Sundrum-type horizon can form.

Similarly, one can take the bulk to be exact Minkowski space.
One can then consider the decay of this Minkowski spacetime to AdS or nothing. This could realize a horizon, with the Witten bubble providing an example.
However, since SUSY-geometries are stable, this requires the existence of non-supersymmetric Minkowski space. The latter is not expected to exist in the string landscape.

As a result, the most interesting situations to study are ETW branes bounding spacetimes without Minkowski or AdS asymptotics. Concretely, one may think of spherical ETW branes bounding ball-shaped universes created from nothing via a `bubble of something'  \cite{Hawking:1998bn,Turok:1998he,Blanco-Pillado:2011fcm,Friedrich:2023tid} process. 
However, we will soon forget about the origin of such universes and focus on their asymptotically late phase and  the conditions for brane-brane horizons to exist.
\subsection{Alternative formulations of a criterion against de-Sitter-like spacetimes}

Before analyzing the Refined Horizon Criterion in detail, we use this subsection to critically discuss possible alternatives.

\paragraph{On a conceivable `Eternal Compact Universe Criterion':}
We could try to forbid
the eternal existence of spatially compact universes. Indeed, when applied to spheres, this would have implications very similar to the Horizon Criterion. This can be understood as follows: Consider the Hartle-Hawking or Linde-Vilenkin creation-from-nothing process
\cite{Vilenkin:1982de,Hartle:1983ai,Linde:1983cm,Vilenkin:1983xq}.
The resulting spatial spheres are driven to collapse by their curvature and to expansion by their vacuum energy. Demanding that their lifetime is finite will then place a abound on how flat asymptotically positive potentials may be, similarly to the Horizon Criterion. However, there is a catch:
As noted long ago \cite{Zeldovich:1984vk,Coule:1999wg,Linde:2004nz}, also spatially flat (e.g.~toroidal) or negatively curved compact universes can form spontaneously.
As there is no curvature effect favoring collapse, they can live eternally, even at vanishing vacuum energy (see however \cite{Bedroya:2025ris}). Hence the eternal existence of compact universes can not be the fundamental issue responsible for the problems with dS.

\paragraph{On a conceivable `Local Asymptotic Acceleration  Criterion':}
One may think that, instead of devising a horizon criterion, one could as well stick with the older and technically more straightforward `no asymptotic acceleration conjecture'. However, it is not clear to us why specifically the acceleration parameter should determine whether a spacetime is allowed or forbidden. Even worse, the definition of the parameter assumes a specific symmetry (FLRW) and a specific parameterization -- this appears to be very far from a fundamental quantum gravity constraint.
This latter issue can of course be easily fixed by devising an appropriate local and covariant criterion.
For example, it has been speculated that the strong energy condition,
\begin{align}
    R_{\mu\nu}v^\mu v^\nu\geq 0\qquad\qquad \text{for all unit timelike vectors}\; v\,\,\,,\label{SEC}
\end{align}
must hold at late times \cite{Rudelius:2021azq}. For FLRW universes, this is equivalent to $\ddot a(\tau)\leq 0$.

As an alternative to \eqref{SEC}, one could imagine that a constraint on the locally defined `relative acceleration of nearby geodesics' $A^\mu$ should be imposed. Explicitly, consider a family of geodesics $x^\mu(t,s)$, parameterized by $s$ and each with proper time $t$. 
We define $s$ such that the deviation vector $X^\mu \equiv \partial_s x^\mu(t,s)$ has unit norm and is orthogonal to the tangent vector $T^\mu \equiv \partial_t x^\mu(t,s)$. 
Then, the geodesic deviation equation relates $A^\mu$ to the Riemann tensor:
\begin{align}
    A^\mu\equiv (T^\nu\nabla_\nu)^2X^\mu = R^\mu_{~\nu\rho\sigma}T^\nu T^\rho X^\sigma\,.\label{intro_geodesic_deviation}
\end{align}
One may thus impose a condition on a local scalar $S(A^\mu)$ built from $A^\mu$ (e.g.~$A_\mu A^\mu$ or $A_\mu X^\mu$). In comparison to \eqref{SEC}, this constrains the Riemann tensor rather than the Ricci tensor and is hence more general.
For an FLRW spacetime with metric \eqref{intro_FLRW_metric}, the geodesic family defined by
\begin{align}
    T^\mu\partial_\mu = \partial_\tau\,,\qquad X^\mu\partial_\mu = \frac{1}{a(\tau)}\partial_y\,,
\end{align}
with $y$ a coordinate of $ds_{d-1}^2$, 
leads to the geodesic deviation equation
\begin{align}
    A^\mu\equiv(T^\nu\nabla_\nu)^2X^\mu = \frac{\ddot a}{a^2}\delta_{y}^{~\mu}\,,\qquad\text{such that}\qquad A_\mu X^\mu=\frac{\ddot a}{a}\,,\qquad A_\mu A^\mu= \left(\frac{\ddot a}{a}\right)^2\,.
\end{align}
A constraint on $A^\mu$ is thus equivalent to a constraint on $\ddot a/a$, similar to \eqref{SEC}.\footnote{For an FLRW spacetime with metric \eqref{intro_FLRW_metric}, $R_{\tau\tau}=-(d-1)\ddot a/a$, such that \eqref{SEC} also directly constrains $\ddot a/a$.}

Requiring \eqref{SEC} or $A_\mu X^\mu\leq 0$ to hold for $\tau>\tau_0$ provides a local condition on the spacetime geometry that forbids eternal accelerated expansion.
However, there exist constructions in string theory where $\ddot a(\tau)$ approaches zero from above for $\tau\to\infty$ [\citealp{Chen:2003dca,Andersson:2006du,Marconnet:2022fmx},~\citealp{Andriot:2023wvg}], such that this condition is too strong.
At best, one could demand that the limiting value of $\ddot a$ at $\tau\to\infty$ is non-positive. Such a statement is then somewhat unsatisfying as there is no quantification of how quickly $R_{\mu\nu}v^\mu v^\nu$ or $S(A^\mu)$ must approach zero. To exclude cosmologies where $\ddot{a}$ decays arbitrarily slowly, \cite{Hebecker:2023qke} proposes
that the equation of state parameter $w$ must asymptote to its critical value or exceed it. But this is of course somewhat involved.

We see that appropriate conditions on $R_{\mu\nu}v^\mu v^\nu$ or $S(A^\mu)$ probably have to involve a nontrivial specification of how fast this quantity must decay asymptotically. This appears ad-hoc and it thus remains unclear whether a convincing constraint involving locally defined quantities can be found. We conclude that the presence of horizons, as suggested in this context in \cite{Hebecker:2023qke, Andriot:2023wvg}, remains in many ways the conceptually simpler and hence hopefully more fundamental property to consider.

\section{ETW Brane with brane-localized Scalar}\label{Sect:Brane_field}
Let us first assume the $d$-dimensional bulk dynamics to be purely gravitational, with vanishing cosmological constant. The bulk is bounded by a timelike ETW brane, with a localized scalar field $\phi$ and potential $V(\phi)$:
\begin{align}
S=\frac{1}{2}M_d^{d-2}\int_{\mathcal{M}} \sqrt{-g}\mathcal{R}_d+\int_{\partial \mathcal{M}} \sqrt{-h}\left(M_d^{d-2}\mathcal{K}_d-\frac{1}{2}M_d^{d-3}h^{ab}\partial_a\phi\partial_b\phi-V(\phi)\right)\,.\label{action_brane_field}
\end{align}
Here $M_d$ denotes the $d$-dimensional Planck mass, $g_{\mu\nu}$ the spacetime metric, $\mathcal{R}_d$ the $d$-dimensional Ricci scalar, $h_{ab}$ the induced metric on the boundary, $\mathcal{K}_d$ the extrinsic curvature scalar of the boundary, and $\mathcal{M}$ the spacetime manifold with boundary $\partial\mathcal{M}$.

We are interested in solutions where a homogeneous and isotropic ETW brane asymptotically expands. This of course includes different sub-cases: The brane could be spatially flat or possess negative or positive curvature. Similarly, the adjacent bulk could also be curved and the curvature could even vary transversely to the wall without reducing the symmetry of the solution. Clearly, all such curvature effects will asymptotically decay and we expect that picking one of the many possibilities outlined above is sufficient to derive the general bound on $V(\phi)$. This expectation is based on the fact that, in the boundary-free situation, it is known that curvature effects are insufficient to induce horizons if the scalar potential is too steep to do so \cite{Andriot:2023wvg}. Nevertheless, this is of course not a proof and further scrutiny may be warranted.
Based on these preliminary remarks, we first study spherical ETW branes bounding a solid cylinder of Minkowski space, with the cross section being a ball.
Based on the intuition developed, we then proceed by studying a setting where the ETW brane is embedded with flat spatial sections.

\subsection{Spherical ETW Brane}\label{Sect:spherical_emb}
The spherical brane bounds a solid cylinder of Minkowski space, with metric
\begin{align}
    ds_g^2=-dt^2+dr^2+r^2d\Omega_{d-2}^2\,.\label{Metric_spherical}
\end{align}
The bulk clearly solves Einstein's equations such that the only non-trivial gravitational dynamics is in the motion of the  spherical ETW brane, located at $r=R(t)$. The induced geometry on the boundary is of FLRW-type and the induced metric may be written as
\begin{align}
    ds_h^2=-d\tau^2+R(\tau)^2d\Omega_{d-2}^2\,,\label{metric_Eigentime}
\end{align}
with $\tau$ the eigentime and $R$ the brane radius.
The eigentime is related to Minkowski time via
\begin{align}
    d\tau = \sqrt{1-(\partial_t R)^2}dt\,,\qquad dt=\sqrt{1+(\partial_\tau R)^2}d\tau\,.\label{Relation_tau_t}
\end{align}
The equations of motion following from the action \eqref{action_brane_field} are\footnote{These equations can be obtained from the Israel junction conditions $M_d^{d-2}(K_{ab}-h_{ab}K)=S_{ab}$ \cite{Israel:1966rt}, where $K_{ab}$ is the extrinsic curvature and $S_{ab}$ is the energy momentum tensor of the fields living on the boundary. The extrinsic curvature is defined as $K_{ab}=\nabla_a n_b$, where $n$ is the outward-pointing unit normal vector of the spacetime boundary.\\
\hspace*{.7cm}Note that our equations differ from much of the brane-cosmology literature (see e.g.~\cite{Kaloper:1999sm, Binetruy:1999ut, Shiromizu:1999wj, Mizuno:2002wa, Maeda:2000mf}) as these papers work locally on the covering space of a $\mathbb{Z}_2$ orbifold. By contrast, we treat the brane locally as a boundary of a half-space. Agreement may be achieved through the identification $\rho_{\rm there}=2\rho_{\rm here}$.}
\begin{align}
&H^2 = \frac{M_d^{-2(d-2)}}{(d-2)^2} \rho^2 - \frac{1}{R^2}\,,\qquad &&H \equiv \frac{\dot R}{R}\,\,\,,\label{Friedmann_brane_field}\\
    &M_d^{d-3}\left(\ddot{\phi}+(d-2)H\dot{\phi}\right) + \partial_\phi V(\phi)=0\,, \qquad &&\rho \equiv \frac{1}{2} M_d^{d-3} \dot{\phi}^2 + V(\phi)\,\,,\label{brane_eom}
\end{align}
and the `dot' stands for $d/d\tau$.
Furthermore, it follows from the derivation of \eqref{Friedmann_brane_field} that the condition $\rho>0$ must hold. Equation \eqref{Friedmann_brane_field} is similar to the Friedmann equation for a closed universe, but with a key difference \cite{Binetruy:1999ut, Shiromizu:1999wj}:
The energy density $\rho$ appears quadratically rather than linearly, implying a substantially different dynamics.

A necessary condition for the existence of brane-brane horizons in the $d$-dimensional spacetime is that the induced, $(d\!-\!1)$-dimensional geometry of the boundary has horizons. The condition for these to occur is well-known: It is given by \eqref{horizon_calc} where the scale factor is identified with $R(\tau)$.
Note that only the asymptotic form of $R(\tau)$ is of relevance.
It follows that if 
\begin{align}
    R(\tau) \sim \tau^\alpha \qquad \text{ for some }\qquad \alpha>1\,,\label{Condition_alpha_1}
\end{align}
the induced geometry has a cosmological horizon. 

In the presently discussed situation, the condition above is not sufficient: Indeed, two points $p,p'$ on the boundary can be causally connected in the full $d$-dimensional spacetime but not in the $(d\!-\!1)$-dimensional geometry of the boundary.  
A sufficient condition for brane-brane horizons to exist is the following: An observer on the ETW brane should not be able to send a signal to the antipodal point of the spherical brane, not even through the bulk. This corresponds to demanding that
\begin{align}
    R(t)-r_\gamma (t)> 0\qquad
    \qquad \mbox{for} \qquad \qquad t\to\infty\,,     
    \label{brane_brane_criterion}
\end{align}
holds for the radial profile $r_\gamma(t)$ of any light-ray sent at some finite time $t=\bar{t}$ from the brane and directed towards the antipodal point. 
For the metric \eqref{Metric_spherical}, $r_\gamma(t)$ is given by
\begin{align}
    r_\gamma(t)=|R(\bar{t})-(t-\bar{t})|\,,
\end{align}
such that the condition \eqref{brane_brane_criterion} can be written as
\begin{align}
    R(t)-t>-\infty\,.
\end{align}
Using the relation \eqref{Relation_tau_t}, one may rewrite this as
\begin{align}
    -\infty < \int^\infty \left(\dot R-\sqrt{1+\dot R^2}\right)d\tau \simeq -\frac{1}{2}\int^\infty\frac{1}{\dot R}d\tau\,,\label{brane_brane_condition_Rdot}
\end{align}
where only the $\tau\to\infty$ region of the integral is important.
As a consequence, we see that if $R(\tau)$ grows faster than
\begin{align}
    R\sim \tau^\alpha\,,\qquad \text{ for some }\qquad \alpha>2\,,\label{brane_brane_horizon_R_tau}
\end{align}
brane-brane horizons exist.

One may think that having found the condition \eqref{brane_brane_horizon_R_tau} rather than \eqref{Condition_alpha_1} is a result of having chosen a spherically symmetric brane embedding.
We are, in the spirit of the swampland program, interested in the `worst-case scenario': We want to potentially rule out potentials which lead to brane-brane horizons in {\it some} geometry, not necessarily in {\it all} geometries. In particular, one may expect \eqref{Condition_alpha_1} to be sufficient if we choose e.g.~a flat brane embedding.
We will study such an embedding momentarily and we find that, nevertheless, \eqref{brane_brane_horizon_R_tau} is the appropriate condition for the existence of a brane-brane horizon.

\subsection{ETW Branes with flat Slicing}\label{Sect:flat_brane}
We believe the spherical case discussed above to be the most intuitive and technically simplest. Nevertheless, as just explained, one could suspect that a flat ETW brane embedding leads to a stronger condition. To clarify this, we build on \cite{Horowitz:2000ds}, but generalize the analysis to arbitrary dimensions. The metric ansatz is
\begin{align}
    \begin{split}
        ds_{\rm g}^2 &= e^{2A(t,z)}(-dt^2+dz^2)+e^{2B(t,z)}ds_{d-2}^2
    =-e^{2A(u,v)}dudv+e^{2B(u,v)}ds_{d-2}^2\,,\label{metric_Hor}
    \end{split}
\end{align}
where $u=t-z$, $v=t+z$ and $ds_{d-2}$ is the spatial metric of the ETW brane. The latter is taken to be localized at $z=0$ and the bulk is the region $z>0$.
The metric \eqref{metric_Hor} follows from considering an FLRW-type brane embedding and choosing conformal gauge in the directions transverse to $ds_{d-2}^2$.
The induced metric on the worldvolume of the ETW brane can be written as
\begin{align}
    ds_{\rm h}^2=-d\tau^2 +R(\tau)^2ds_{d-2}^2\,,
\end{align}
with the eigentime $\tau$ and the scale factor $R(\tau)$ given by
\begin{align}
    d\tau=e^{A(t,z=0)}dt\,,\qquad R(\tau)=e^{B(t(\tau),z=0)}\,.\label{eigentime_Hor}
\end{align}
We consider the spatial geometry $ds_{d-2}^2$ of the ETW brane to be flat euclidean space. It can either be infinitely extended or compact, e.g.~toroidal.
The Einstein equations following from the above metric ansatz are
\begin{align}
    \partial_u\partial_v B &+ (d-2)\partial_uB\partial_vB=0\,,\\
    \partial_uA\partial_uB&-\frac{1}{2}\left((\partial_uB)^2+\partial_u^2B\right)=0\,,\label{Einstein_nontrivial}\\
    \partial_vA\partial_vB&-\frac{1}{2}\left((\partial_vB)^2+\partial_v^2B\right)=0\,,\\
    \partial_uA\partial_vA&+(d-3)\partial_u\partial_vB+\frac{1}{2}(d-2)(d-3)\partial_uB\partial_vB=0\,,
\end{align}
and the Israel junction conditions take the form\footnote{The unit normal vector is given by $n=-e^{-A}\partial_z$.}
\begin{align}
    -M_d^{d-2}(d-2)e^{-A(t,0)}\partial_zB|_{z=0}=\rho\,,\label{Israel_1}\\
    M_d^{d-2}e^{-A(t,0)}((d-3)\partial_zB+\partial_zA)|_{z=0}=p\,,\label{Israel_2}
\end{align}
with $\rho$ and $p$ being the energy density and pressure on the ETW brane induced by the field $\phi$.
We now make an important assumption:
We consider $A$ and $B$ to depend on $u=t-z$ only.
One may understand this requirement as choosing the boundary conditions such that no incoming flux of energy hits the ETW brane.
Furthermore, this assumption ensures that the bulk geometry is Riemann-flat.
It follows that there remains a single nontrivial Einstein equation \eqref{Einstein_nontrivial} to be solved.
It can be integrated to yield\footnote{The integration constant can be absorbed into a redefinition of $u$.}
\begin{align}
    A(u)=\frac{1}{2}B(u)+\frac{1}{2}\log(\partial_uB)\,.\label{sol_Einstein_brane}
\end{align}
Furthermore, using the fact that $\partial_zB=-\partial_tB$, we can rewrite the Israel condition \eqref{Israel_1} as
\begin{align}
    \frac{M_d^{-(d-2)}\rho}{d-2}&=e^{-A}\partial_tB=\partial_\tau B=\frac{\partial_\tau e^B}{e^B}=\frac{\partial_\tau R}{R}\,.
\end{align}
We thus find, in accordance with \cite{Shiromizu:1999wj,Maartens:2003tw},
\begin{align}
    H^2=\frac{M_d^{-2(d-2)}}{(d-2)^2}\rho^2\,,\qquad H=\frac{\dot R}{R}=\frac{\partial_\tau R}{R}\,,\label{Friedmann_brane_flat}
\end{align}
the Friedmann equation of type \eqref{Friedmann_brane_field}.
Because of the flat brane slicing the curvature term $R^{-2}$ is absent. Equation \eqref{Friedmann_brane_flat} and the scalar field equation \eqref{brane_eom} determine the evolution of the system. 
Our assumption that $A$ and $B$ depend only on $u=t-z$ is related to setting the projected 5d Weyl tensor $E_{ab}$ to zero, see e.g.~\cite{Shiromizu:1999wj, Maartens:2003tw}. We note that the second Israel condition \eqref{Israel_2} is a consequence of \eqref{Israel_1} and the the scalar field equation \eqref{brane_eom}.

Since the brane geometry is spatially flat, one might think that sending light-rays through the bulk is not advantageous in terms of establishing causal contact between brane points. If so, horizons in the induced geometry (i.e.~condition~\eqref{Condition_alpha_1}) would imply brane-brane horizons in the full geometry. However, this intuition turns out to fail, as we now demonstrate. 

In App.~\ref{App:A_B_u}, it is shown that for a profile $R\sim\tau^\alpha$ with $\alpha>1$, $\tau$ is given by the following function of $t$
\begin{align}
    \tau=\left[\alpha(2-\alpha)(t-u_0)\right]^{\frac{1}{2-\alpha}}\,.
\end{align}
Importantly, this shows that for $\alpha>2$, the coordinate range of $t$ at the ETW brane is bounded by some constant $u_0$.
In other words, $\tau\to\infty$ corresponds to $t\to u_0$.
By contrast, for $1<\alpha<2$, $\tau\to\infty$ corresponds to $t\to\infty$.
Furthermore, it is shown in App.~\ref{App:A_B_u} that the functions $e^{A(u)}$ and $e^{B(u)}$ behave as
\begin{align}
\begin{split}
    e^{A(u)}&=\alpha\left[\alpha(2-\alpha)(u-u_0)\right]^{\frac{\alpha-1}{2-\alpha}}\,,\\
    e^{B(u)}&=\left[\alpha(2-\alpha)(u-u_0)\right]^{\frac{\alpha}{2-\alpha}}\,,\label{e_A_B}
\end{split}
\end{align}
where the relevant parameter range is $u\gtrless u_0$ for $\alpha \lessgtr 2$. Following these preliminaries, we now show that brane-brane horizons only appear for $\alpha>2$.
Let $x$ denote a coordinate of $ds_{d-2}^2$ and consider two observers on the ETW brane located at $x=0$ and $x=x_0$. 
They are in causal contact with each other if, for all $\Bar{t}$, the observer at $x=0$ can send a signal starting at $t=\Bar{t}$ to the observer at $x=x_0$. 
We consider a light-like trajectory $x(t),z(t)$ in the $x,z$ plane starting at $t=\bar{t}$ at $x(\Bar{t})=0$ and $z(\Bar{t})=0$. 
The condition of being light-like, $ds_{\rm g}^2=0$, implies
\begin{align}
    e^{2A(t-z(t))}(1-z'(t)^2)=e^{2B(t-z(t))}x'^2\,,\label{ll_traj}
\end{align}
where a prime denotes a derivative with respect to $t$.
The trajectory $x(t)$ is then determined by the function $z(t)$ via
\begin{align}
    x(t)=\int_{\Bar{t}}^t e^{(A-B)(\tilde{t}-z(\tilde{t}))}\sqrt{1-z'(\tilde{t})}\;d\tilde{t}\,.\label{x_t}
\end{align}

Let us first consider the case $\alpha>2$.  Note that the integrand of \eqref{x_t} approaches zero as $\bar{t}$ approaches its maximal possible value $u_0$. Thus, when 
when choosing $\Bar{t}$ close to $u_0$, the integral becomes arbitrary small independently of the trajectory $z(t)$.
As a result, there exists a $\Bar{t}$ such that for $t>\Bar{t}$, the observer at $x=0$ can no longer send a signal to the observer at $x=x_0$.
This implies that brane-brane horizons exist. 

Let us now consider the case $1<\alpha<2$, for which $t$ is unbounded from above. 
In terms of the variable $u(t)=t-z(t)$, the integral \eqref{x_t} reads
\begin{align}
    x(t)=\int_{\Bar{t}}^{u(t)} e^{(A-B)(u)}\frac{\sqrt{(1-z')(1+z')}}{1-z'}du\,.
\end{align}
For $z'>0$, it follows that (remember $1-z'=u'$)
\begin{align}
    x(t)= \int_{\Bar{t}}^{u(t)} e^{(A-B)(u)}\sqrt{\frac{2}{u'}-1}\,du\,.
\end{align}
Let us now consider a trajectory $z(t)=z_*(t)$, or equivalently $u(t)=u_*(t)=t-z_*(t)$, defined by
\begin{align}
    e^{(A-B)(u_*)}\sqrt{\frac{2}{u'_*}-1}=1\,.\label{u_star_diff}
\end{align}
For such a trajectory, one finds
\begin{align}
    x_*(t)= u_*(t)-\Bar{t}\,.\label{x_t_const}
\end{align}
Inserting the profile \eqref{e_A_B} into \eqref{u_star_diff} leads to
\begin{align}
    t=\frac{1}{2}\left(u_*(t)+c_\alpha (u_*(t))^{\frac{4-\alpha}{2-\alpha}}\right)+\text{const}\,,
\end{align}
with $c_\alpha>0$ an $\alpha$-dependent constant, such that $u_*(t\to\infty)\to\infty$.
From \eqref{x_t_const}, we see that there must then exist a $t_*$ such that $x_*(t_*)=x_0$. 
We can then consider the trajectory
\begin{align}
	z(t)=\begin{cases}
		z_*(t) & \Bar{t}<t<t_*\,,\\
		z_*(t_*)-(t-t_*) & t_*<t<t_*+z_*(t_*)
	\end{cases}\,,
\end{align}
which is a solution to \eqref{ll_traj} and satisfies $x(\Bar{t})=0$, $z(\bar{t})=0$, as well as $x(t_*+z_*(t_*))=x_0$, $z(t_*+z_*(t_*))=0$.
Since $x_0$ and $\Bar{t}$ are arbitrary, it follows that brane-brane horizons do not exist, as the above trajectory corresponds to a signal sent from the observer at $x=0$ to the observer at $x=x_0$.

In summary, we found that brane-brane horizons exist if 
\begin{align}
    R\sim\tau^\alpha\,\qquad\text{with}\qquad \alpha>2\,.\label{R_alpha_g_2_flat}
\end{align} 
This result, now derived a flat brane embedding of type \eqref{metric_Hor}, is in agreement with condition \eqref{brane_brane_horizon_R_tau}
that we obtained previously for the spherically embedded brane. One may find this surprising since our earlier derivation heavily relied on the spherical geometry of the ETW brane.
A careful study of the geometry \eqref{metric_Hor} allows for an explanation:
As the bulk spacetime is Riemann-flat, there must exist  coordinates $T,w,X_2,...X_{d-2}$, in which the metric takes the standard form
\begin{align}
    ds^2=-dT^2+dw^2+dX_2^2+...+dX_{d-2}^2\,.
\end{align}
The corresponding coordinate transformation is derived in App.~\ref{App:Embedding} for the case $1<\alpha<2$. There, it is also shown that the ETW brane embedding is specified by the equation 
\begin{align}
    X_1^2+...+X_{d-2}^2+w^2=T^2-\frac{(T-w)^\frac{2}{\alpha}}{2-\alpha}\,.
\end{align}
We see that, at late Minkowski time $T$, the spatial geometry of the ETW brane is spherical and the brane expands with the speed of light.
The bulk, which used to correspond to $z>0$, now maps to a cylindrical region bounded by a spatial sphere at every moment of time. This is very similar to the spherically embedded ETW brane discussed before. The situation is similar to the well known case of de Sitter space, which allows for both a closed and a flat slicing.

Finally we recall that, in the geometry \eqref{metric_Hor}, the condition for the existence of brane-brane horizons may be formulated as follows: The limit $\tau\to\infty$ on the ETW brane should correspond to a finite limit, $t\to u_0$, in terms of the bulk time $t$. In App.~\ref{App:Critical_brane}, we show that this latter condition is equivalent to \eqref{brane_brane_condition_Rdot}, which we also found when studying a spherical brane embedding.
We may thus conjecture that \eqref{brane_brane_condition_Rdot} is the general condition for the existence of brane-brane horizons.
However, we can at present not exclude the existence of geometries in which the condition for brane-brane horizons to occur is given by \eqref{horizon_calc} (with $a$ replaced by $R$) rather than by \eqref{brane_brane_condition_Rdot}.

\subsection{Implications for the Potential}
Let us now study the conditions on the potential $V(\phi)$ following from  \eqref{brane_brane_horizon_R_tau}.

Before doing so, let us recall some obvious facts:  If $V(\phi)$ has a positive-energy minimum or if it approaches a positive value at $\phi\to \infty$, horizons always appear: The ETW brane becomes asymptotically a constant-tension object, $H$ tends to a constant and the induced geometry approaches 3d de Sitter. The solution for $R(\tau)$ is then given by $R\sim H^{-1}\exp(H\tau)$, ensuring the presence of brane-brane horizons.
Thus, we may restrict our attention to the non-trivial case where $V(\phi)>0$ and $V(\phi)\to 0$ monotonically as $\phi\to \infty$. 

The first key observation is that the critical power-law behavior
\begin{align}
    R(\tau)\sim \tau^\alpha\label{R_tau}
\end{align}
arises if the potential takes the form 
\begin{align}
    V(\phi)=M_d^{d-1}\frac{c^2}{\phi^2}\,.\label{brane_field_critical_potential}
\end{align}
Here, $c$ is related to $\alpha$ by
\begin{align}
    c=\sqrt{4(d-2) \alpha - 2}\,.\label{relation_c_alpha}
\end{align}
This is known in the case $d=5$, where the corresponding scaling solutions have been studied in the context of brane cosmologies \cite{Mizuno:2002wa} (see also \cite{Maeda:2000mf, Mizuno:2004xj}).

To obtain the above relation, one may insert the ansatz \eqref{R_tau} into \eqref{Friedmann_brane_flat}:\footnote{For the spherical embedding of Sect.~\ref{Sect:spherical_emb}, the Friedmann-type equation \eqref{Friedmann_brane_field} contains an extra $1/R^2$ term. However, this term decays decays as $1/\tau^{2\alpha}$ and hence becomes irrelevant at large $\tau$ if $\alpha>1$.}
\begin{align}
    \frac{\alpha}{\tau}=H=\frac{M_d^{-(d-2)}}{(d-2)}\rho\,.\label{rho_tau}
\end{align}
Using \eqref{Friedmann_brane_field},\eqref{Friedmann_brane_flat} and \eqref{rho_tau}, one derives the continuity equation in the form
\begin{align}
    \dot\rho = \dot\phi(M_d^{d-3}\ddot\phi+\partial_\phi V)=-(d-2)M_d^{d-3}H\dot\phi^2=-(d-2)M_d^{d-3}\frac{\alpha}{\tau}\dot\phi^2\,.\label{continuity}
\end{align}
Taking a derivative of \eqref{rho_tau} then results in 
\begin{align}
    \frac{\alpha}{\tau^2}=\frac{\alpha}{\tau}\dot\phi^2\,,
\end{align}
from which it follows that
\begin{align}
    \dot\phi = 1/\sqrt{\tau}\qquad\qquad \mbox{and}\qquad\qquad \phi \simeq 2\sqrt{\tau}
\end{align}
at $\tau\to\infty$.
Inserting this result back into \eqref{rho_tau} and using the definition of $\rho$ from \eqref{Friedmann_brane_field} yields the relation \eqref{relation_c_alpha}.
We thus conclude that \eqref{brane_brane_horizon_R_tau} and hence the existence of brane-brane horizons is guaranteed for
\begin{align}
    c>c_{\rm crit}\equiv \sqrt{2(4d-9)}\,.
    \label{ccr}
\end{align}
Meanwhile, as discussed around \eqref{Condition_alpha_1}, the induced geometry has horizons for $\alpha>1$ which, by using \eqref{relation_c_alpha}, turns into 
\begin{align}
    c>\sqrt{4d-10}\,.
\end{align}

The result \eqref{ccr} can be generalized to a wider class of potentials. To do so, note that the quantity $d(V^{-1/2}(\phi))/d\phi$ represents one possible measure of how fast $V$ approaches zero at large $\phi$. Applied to potentials of the form $V\sim 1/\phi^a$, this quantity diverges for $a>2$, goes to zero for $a<2$, and extracts the value of the constant prefactor for $a=2$. It is then intuitive that brane-brane horizons exist if
\begin{align}
    \frac{d}{d\phi} \sqrt{\frac{M_d^{d-1}}{V(\phi)}} = \sqrt{M_d^{d-1}}\frac{\partial_\phi V}{V^{3/2}} < \frac{1}{c_{\rm crit}} - \epsilon\qquad\text{ for} \qquad \phi>\phi_0\label{Class_brane_field}
\end{align}
holds for some $\epsilon > 0$ and $\phi_0<\infty$. They do not exist if
\begin{align}
    \frac{d}{d\phi} \sqrt{\frac{M_d^{d-1}}{V(\phi)}} \geq \frac{1}{c_{\rm crit}}\qquad\text{ for} \qquad \phi>\phi_0\,.
    \label{cvc}
\end{align}
More details are provided in App.~\ref{App:Critical_brane}.
Furthermore, as explained in App.~\ref{App:H_R_V_phi}, given any positive profile $H(\tau)$ for the expansion rate, there exists a potential $V(\phi)$ that realizes such a cosmology. 
One may now study profiles $H(\tau)$ that barely realize/not realize brane-brane horizons and then compute the corresponding potential.

Equations \eqref{ccr}--\eqref{cvc} together with the definition \eqref{brane_field_critical_potential} are the first consequences of the Refined Horizon Criterion.

One may now be interested in situations when $d(V^{-1/2}(\phi))/d\phi$ approaches $1/c_{\rm crit}$ from below for $\phi\to\infty$. 
In App.~\ref{App:Critical_brane}, we consider potentials that are to leading order of the form \eqref{brane_field_critical_potential} with $c=c_{\rm crit}$ and study conditions on corrections to it in order for brane-brane horizons to exist. 
In particular, we find that potentials of the form 
\begin{align}
    V(\phi)=\frac{c_{\rm crit}^2}{\phi^2}\left(1+\frac{\eta}{\ln(\phi)}\right)
\end{align}
lead to brane-brane horizons if
\begin{align}
    \eta > \eta_{\rm crit}=\frac{d-2}{c_{\rm crit}^2}=\frac{d-2}{2(4d-9)}\,.
\end{align}

As a final comment, let us briefly think about explicit stringy examples for brane-localized scalars: One class of models are those of `Verlinde type' \cite{Verlinde:1999fy} (mentioned in the Introduction), where a 6d compact space realizes the 4d ETW brane (concretely a $T^6/\mathbbm{Z}_2$) and a 5d compact space (concretely a fluxed $S^5$) realizes the bulk by dimensional reduction. The brane scalar would be the volume of the $T^6/\mathbbm{Z}_2$. 
Crucially, if this volume rolls to infinity, higher-dimensional brane-localized operators (in our case a 4d Einstein-Hilbert term) may come to dominate. This situation could arise fairly generally since infinite distances are almost always related to some type decompactification \cite{Lee:2019wij}.
Thus, more work is required to make contact between our simple analysis and string models. Alternatively, if both bulk and brane decompactify or roll to weak coupling together, the above issue is avoided. Thus, we now turn to the case with a bulk scalar field which, as we just argued, may be the more interesting one from the string-theory perspective.

\section{ETW brane coupled to massless Bulk Scalar}\label{Sect:Bulk_field}
Let us now proceed to the case of a bulk scalar field with a brane-localized potential:
\begin{align}
    S=\frac{1}{2}M_d^{d-2}\int_{\mathcal{M}} \sqrt{-g}\left(\mathcal{R}_d-g^{\mu\nu}\partial_\mu\phi\partial_\nu\phi\right)+\int_{\partial \mathcal{M}} \sqrt{-h}\left(M_d^{d-2}\mathcal{K}_d-V(\phi)\right)\,.\label{action_bulk_field}
\end{align}
As already explained in the Introduction, one may think of $\phi$ as a bulk modulus. The brane generically has less supersymmetry and it is hence natural to expect that it provides a non-trivial potential depending on $\phi$. 
We do not include a brane-localized kinetic term for $\phi$: While such a term is generically present, its effects are sub-leading \cite{Daniel_Master}.\footnote{On the one hand, brane-parallel spatial gradients decay quickly as the brane expands. On the other hand, the kinetic energy $\dot{\phi}^2$ receives contributions from the brane and from the adjacent bulk region, with the latter dominating. This can be traced to a $1/M_P$ suppression arising on dimensional grounds \cite{Daniel_Master}.}

\subsection{The Equations of Motion}\label{Sect:Bulk_EoM}
The dynamics is governed by the bulk Einstein equations, the Israel junction conditions, the bulk equation of motion for $\phi$, and its boundary conditions at the ETW brane:
\begin{eqnarray}
    M_d^{d-2}G_{\mu\nu}&=&T_{\mu\nu}\equiv -\frac{M_d^{d-2}}{2}g_{\mu\nu}(\partial\phi)^2+M_d^{d-2}\partial_\mu\phi\partial_\nu\phi\,,\label{Einstein_Eq}\\
    M_d^{d-2}(\mathcal{K}_{d,\tau\tau}+\mathcal{K}_d) &=&  \rho=V(\phi)\,\label{Friedmann_bulk_field}\\
    \phantom{\Big[} 0&=&\square_{\rm g} \phi\,,\label{Phi_EoM_bulk}\\
    0&=&\left(\left.-M_d^{d-2}n^\mu\partial_\mu\phi-\partial_\phi V\right)\right|_{\rm ETW}\,,\label{phi_bdry_condition_bulk}
\end{eqnarray}
where $\mathcal{K}_{d,\tau\tau}$ is the $\tau\tau$-component of the extrinsic curvature tensor. As before, $\tau$ labels the eigentime of the ETW brane such that $h_{\tau\tau}=-1$. 
The boundary condition \eqref{phi_bdry_condition_bulk} is the straightforward generalization of the familiar Neumann boundary condition to the case with a boundary-localized potential, see e.g.~\cite{Chamblin:1999ya}. 
In the equations above,  $n^\mu$ is the (outwards-pointing) unit normal vector of the spacetime boundary and $|_{\rm ETW}$ stands for evaluation at the ETW brane. As before, we focus on positive potentials that approach zero monotonically as $\phi\to\infty$.

We again build on \cite{Horowitz:2000ds} and consider a metric ansatz of the form \eqref{metric_Hor}. As before, we assume the functions $A$ and $B$ as well as the scalar field $\phi$ to depend on $u=t-z$ only. 
The single remaining nontrivial Einstein equation now takes the form
\begin{align}
\partial_uA\partial_uB&-\frac{1}{2}\left((\partial_uB)^2+\partial_u^2B\right)=\frac{(\partial_u\phi)^2}{2(d-2)}\,,\label{Einstein_nontrivial_bulk}
\end{align}
and the boundary condition for $\phi$, Eq.~\eqref{phi_bdry_condition_bulk}, becomes
\begin{align}
    M_d^{d-2}e^{-A(t,0)}\partial_z\phi|_{z=0} = \partial_\phi V(\phi)\,.\label{bdry_phi_z}
\end{align}
The bulk field equation \eqref{Phi_EoM_bulk} is automatically solved by the ansatz $\phi=\phi(u)$.
In terms of the eigentime of the brane \eqref{eigentime_Hor}, the boundary equations, i.e.~Eqs.~\eqref{Friedmann_brane_flat} and \eqref{bdry_phi_z}, read
\begin{align}
    H^2=\frac{M_d^{-2(d-2)}}{(d-2)^2}\rho^2=\frac{M_d^{-2(d-2)}}{(d-2)^2}V^2\,,\qquad \dot\phi = - \frac{\partial_\phi V}{M_d^{d-2}}\,.\label{bdry_eqs_Hor}
\end{align}
The second equation can be interpreted as saying that, at the location of the brane, the field $\phi$ is subject to friction force.
In App.~\ref{App:A_B_u_bulk}, it is shown that the above equations imply
\begin{align}
    A(u)=B(u)\,,\label{A_eq_B}
\end{align}
where the difference to \eqref{sol_Einstein_brane} is caused by the backreaction of $\phi$ in the spacetime bulk.
The spacetime is thus conformally flat, 
\begin{align}
    ds^2=e^{2B(u)}(-dudv+ds_{d-2}^2)\,.
\end{align}
In such a geometry, brane-brane horizons exist if and only if the coordinate $t=(u+v)/2$ at the ETW brane is bounded from above. Let us explain why this is the case: If $t$ is unbounded from above, the induced geometry does not have cosmological horizons. This follows from checking the condition \eqref{horizon_calc}:
\begin{align}
    \int^\infty\frac{d\tau}{R(\tau)}=\int^\infty dt=\infty\,,   
\end{align}
where we used \eqref{eigentime_Hor} and \eqref{A_eq_B}.
If $t$ at the ETW brane is bounded from above by $u_0$, then 
\begin{align}
    \int^\infty\frac{d\tau}{R(\tau)}=\int^{u_0} dt<\infty\,,   
\end{align}
and the induced geometry has cosmological horizons. 
Furthermore, a signal sent from the ETW brane at time $u_0-\Delta$ can at most travel a coordinate distance $\Delta$ in the metric $ds_{d-2}^2$ until reaching $t=u_0$.
By considering small $\Delta$, it follows that an observer on the ETW brane is out of causal contact with some regions of the ETW brane and hence brane-brane horizons exist.

It is shown in App.~\ref{App:A_B_u} that for solutions of the form $R\sim\tau^\alpha$, the functions $A(u)$ and $B(u)$ take the form
\begin{align}
    e^{A(u)}=e^{B(u)}\sim\left[-(\alpha-1)(u-u_0)\right]^{-\frac{\alpha}{\alpha-1}}\,,
\end{align}
with the relevant parameter range now being $u\gtrless u_0$ for $\alpha \lessgtr 1$.
We thus find that brane-brane horizons exist if
\begin{align}
    R\sim\tau^\alpha\,,\qquad\text{with}\qquad \alpha > 1\,.\label{R_alpha_1_bulk_field}
\end{align}
The difference to condition \eqref{R_alpha_g_2_flat} is caused by the backreaction of $\phi$ in the bulk.
It is clear from the discussion above (see also App.~\ref{App:Critical_bulk}) that the condition for the time $t$ at the ETW brane to be bounded from above is equivalent to \eqref{horizon_calc}, with $a$ replaced by $R$.
Thus, brane-brane horizons arise if the induced geometry has cosmological horizons.

It would be interesting to consider scenarios where $\phi$ has a nontrivial potential in the bulk, in addition to the brane potential. The modified backreaction of $\phi$ in the bulk may change the condition \eqref{R_alpha_1_bulk_field} for brane-brane horizons to occur.

\subsection{Implications for the Potential}
Let us now study, for the modulus case, which brane potentials lead to an expansion of the form $R\sim\tau^\alpha$.
From Eq.~\eqref{bdry_eqs_Hor}, it follows that
\begin{align}
    V(\phi(\tau))\sim (d-2)M_d^{d-2}\frac{\alpha}{\tau}\,,\qquad \dot\phi = - \frac{\partial\tau}{\partial\phi}\frac{\partial_\tau V}{M_d^{d-2}}\sim\frac{1}{\dot\phi}(d-2)\frac{\alpha}{\tau^2}\,.\label{V_tau_bulk}
\end{align}
The second relation implies that
\begin{align}
    \phi\sim\sqrt{(d-2)\alpha}\log(\frac{V_0}{(d-2)M_d^{d-2}\alpha}\tau)\,,\label{phi_tau_bulk}
\end{align}
where we have introduced a dimensionful parameter $V_0$. 
Finally, inverting \eqref{phi_tau_bulk} and plugging the result into \eqref{V_tau_bulk} yields
\begin{align}
    V(\phi)\sim V_0\exp(-\lambda \phi)\,,\qquad \text{with}\qquad\lambda = \frac{1}{\sqrt{(d-2)\alpha}}\,.\label{potential_bulk_field}
\end{align}
From \eqref{R_alpha_1_bulk_field}, it follows that a sufficient condition for brane-brane horizons to exist is given by
\begin{align}
    \lambda <\lambda_{\rm crit}\equiv \frac{1}{\sqrt{d-2}}\,.\label{Horizon_critereon_bulk_field}
\end{align}

The form of the critical potential, an exponential in $\phi$, is similar to the one found when studying cosmological horizons for universes without boundaries. 
However, the critical parameter in \eqref{Horizon_critereon_bulk_field} differs from that in \eqref{horizon_criterion_bulk} by a factor $1/2$.

Moreover, we can compare the critical potential of \eqref{potential_bulk_field} with the analogous result in \eqref{brane_field_critical_potential}, obtained for a field $\phi$ living solely on the ETW brane.  We observe that potentials realizing brane-brane horizons in the case of a bulk scalar field are allowed to be much steeper -- exponential rather than power-like. 

The result \eqref{Horizon_critereon_bulk_field} can be generalized to a wider class of potentials. As in the discussion around \eqref{Class_brane_field}, it is useful to select an expression quantifying the steepness of the potential. In the present case, an appropriate choice is $-\partial_\phi V/V$.
For potentials of the form \eqref{potential_bulk_field}, this is simply $\lambda$.
It is thus intuitive (see also App.~\ref{App:Critical_bulk}) that for a given $V(\phi)$, brane-brane horizons exist if
\begin{align}
    -\frac{\partial_\phi V}{V}<\lambda_{\rm crit}-\epsilon\,,\qquad\text{for all}\qquad \phi>\phi_0\,,\label{Horizon_critereon_bulk_field_general}
\end{align}
for some constants $\phi_0$ and $\epsilon>0$.
They do not exist in case
\begin{align}
    -\frac{\partial_\phi V}{V}\geq\lambda_{\rm crit}\,,\qquad \text{for} \qquad \phi>\phi_0\,.\label{Horizon_critereon_bulk_field_general_no_horizon}
\end{align}
As before, given any positive profile of the expansion rate $H(\tau)$, there exists a corresponding potential that realizes this cosmology, as explained in App.~\ref{App:H_R_V_phi_bulk}.

Furthermore, it is possible to consider potentials for which the decisive quantity $-\partial_\phi V/V$ asymptotes to 
$\lambda_{\rm crit}$ and study which of them induce brane-brane horizons. Concretely, it is shown in App.~\ref{App:Critical_bulk} that potentials of the form
\begin{align}
    V=V_0\phi^\beta\exp(-\lambda_{\rm crit}\phi)\,\qquad \text{with}\qquad \beta>\frac{1}{2}
    \label{cpwb}
\end{align}
lead to brane-brane horizons. If $\beta\leq 1/2$, no brane-brane horizons arise.

The corresponding result for universes without boundary is derived in App.~\ref{App:Critical_closed}.
This represents a generalization of \eqref{horizon_criterion_bulk}. We find that potentials of the form 
\begin{align}
    V=V_0\phi^\beta \exp(-\gamma_{\rm crit}\phi)\,\qquad \text{with}\qquad \beta >1
    \label{cpwob}
\end{align}
lead to cosmological horizons. If $\beta\leq 1$, no horizons arise.

It is remarkable that, even at sub-leading order, the critical potentials in \eqref{cpwob} are the square of critical potentials in \eqref{cpwb}. We have so far not been able to find a conceptual reason for this intriguing coincidence.

\section{Conclusions}
In this paper, we extended the Horizon Criterion, which identifies spacetimes with cosmological horizons as being problematic in quantum gravity,
to spacetimes with boundaries.
A spacetime $\mathcal{M}$ is said to have cosmological horizons if there exists an inertial observer that is causally disconnected from some parts of $\mathcal{M}$. If $\mathcal{M}$ 
has a dynamical boundary, i.e.~an end-of-the-world (ETW) brane, such an 
observer can be bound to that brane.
The most naive extension of the Horizon Criterion would be to demand that all inertial observers, including both bulk and brane observers, do not experience a cosmological horizon. 
However, one quickly realizes that observers living on a Randall-Sundrum (RS) UV brane do see a cosmological horizon (cf.~Fig.~\ref{Fig:RS_Brane}), as there are points in the AdS bulk spacetime that are causally disconnected from the ETW brane.
Now, there is no fundamental reason to believe that the RS model is inconsistent in quantum gravity and, what is more, there exists an embedding in string theory \cite{Verlinde:1999fy}. It would be interesting to study the RS horizon from a brane-EFT or holographic point of view.
In any case, the Horizon Criterion can not be applied in its most naive form.

One possible conclusion from this observation would be that cosmological horizons simply do not represent a fundamental problem in quantum gravity.
One may also argue against horizon-type criteria because the definition of a horizon is intrinsically non-local. As a result, a certain amount of arbitrariness is introduced since one must decide to which set of observers it applies.
This takes us back to the originally discussed possibility that the problem is rooted in asymptotic acceleration. However, that appears unlikely since negatively curved universes accelerate even with steep potentials. Thus, maybe one should be more optimistic that de Sitter and similar cosmologies are not in the Swampland after all, or one must look for a better criterion excluding them. 

Alternatively, one would have to refine the Horizon Criterion. We argued that an appropriate generalization is to demand that any two observers living on the same ETW brane should be in causal contact with each other. To be more precise, we introduced the notion of a \textit{brane-brane horizon}. By this we mean the intersection of the cosmological horizon of a brane-bound observer with the brane itself. The Refined Horizon Criterion then says that no brane-brane horizons should exist.
We analyzed the consequences following from such a criterion for ETW branes whose action includes a brane localized potential $V(\phi)$, with $\phi$
a scalar field. For ETW branes bounding Anti-de Sitter (AdS) space, the corresponding conditions can be easily inferred using AdS/CFT correspondence. Hence, we focused on the more interesting case where a $(d\!-\!1)$-dimensional ETW brane bounds a $d$-dimensional geometry with vanishing cosmological constant.
Brane-localized potentials with a positive energy minimum or approaching a positive value asymptotically trivially realize horizons and are hence excluded. Our main object of interest were hence brane-localized potentials $V(\phi)$ monotonically approaching zero from above at $\phi\to\infty$.

We started with the analysis of models where $\phi$ is localized on the ETW brane. The maybe simplest spatial geometry to consider is one where a spherical ETW brane bounds a ball of bulk space. Dynamically, the ETW-brane grows while the inner region is unperturbed Minkowski space.
We found that the condition for the presence of brane-brane horizons reads
\begin{align}
    \int^\infty\frac{1}{\partial_\tau R(\tau)}d\tau < \infty\,,\label{conc_R_dot_tau}
\end{align}
where $R(\tau)$ is the scale factor of the induced FLRW-geometry on the ETW brane and $\tau$ is the eigentime of the brane. 
This condition differs from 
\begin{align}
    \int^\infty \frac{1}{R(\tau)}d\tau<\infty\,,
\end{align}
the familiar condition for cosmological horizons to arise in FLRW spacetimes. Technically, the conditions differ because, in the brane-case, two observers on the ETW brane can be in causal contact via bulk trajectories even though they are separated by a horizon in the induced brane geometry. One might think that this is an artifact of choosing a spherical embedding of the ETW brane, but we showed that the same condition \eqref{conc_R_dot_tau} follows when considering a metric ansatz in which the embedding of the ETW brane is spatially flat. 
A more detailed analysis of the geometry revealed that the resulting induced geometry also allows for a spherical slicing. We expect this to be a general feature, i.e., we expect that an ETW brane with positive energy bounding a bulk with zero vacuum energy always allow for a closed slicing.  As a result, we consider \eqref{conc_R_dot_tau} to be the general condition which arises from demanding the absence of brane-brane horizons.

The evolution of $R(\tau)$ is governed by a Friedmann-type equation that differs from the ordinary Friedmann equation by the energy density appearing quadratically rather than linearly. 
We showed that brane-cosmologies satisfying \eqref{conc_R_dot_tau} are obtained for potentials $V(\phi)$ that satisfy 
\begin{align}
    \frac{d}{d\phi} \sqrt{\frac{M_d^{d-1}}{V(\phi)}} < \frac{1}{c_{\rm crit}} - \epsilon\label{conclusion_brane_field}\,,\qquad c_{\rm crit}=\sqrt{2(4d-9)}\,,
\end{align}
for $\phi>\phi_0$ with arbitrary constants $\phi_0$ and $\epsilon>0$. By contrast, if
\begin{align}
    \frac{d}{d\phi} \sqrt{\frac{M_d^{d-1}}{V(\phi)}} \geq \frac{1}{c_{\rm crit}} \label{conclusion_brane_field_2}\,
\end{align}
for $\phi>\phi_0$, brane-brane horizons do not arise. The `measure of steepness' of $V$ appearing on the l.h.~side of \eqref{conclusion_brane_field} and \eqref{conclusion_brane_field_2} is constant for potentials of the form
\begin{align}
    V(\phi)=M_d^{d-1}\frac{c^2}{\phi^2}\,.\label{conc_V_brane}
\end{align}
In this sense, such potentials represent the critical case. They induce brane-brane horizons if $c>c_{\rm crit}$ and avoid them if 
$c\leq c_{\rm crit}$. 
By contrast, the induced geometry on the boundary has cosmological horizons for $c>\sqrt{4d-10}$.
Moreover, using an autonomous system analysis we studied the case of critical $c$ with asymptotically subleading terms included:
\begin{align}
V(\phi)=M_d^{d-1}\frac{c_{\rm crit}^2}{\phi^2}\left(1+\frac{\eta}{\log(\phi)}\right)\,.\label{conc_V_eta}
\end{align}
We found that brane-brane horizons arise if
\begin{align}
    \eta>\eta_{\rm crit}=\frac{d-2}{2(4d-9)}\,.
\end{align}

We also analyzed the scenario where $\phi$ is a bulk field with exactly vanishing bulk potential, subject to an ETW-brane-localized potential $V(\phi)$.
In this case, due to the backreaction of $\phi$ in the bulk, the condition for horizons changes w.r.t.~the brane-scalar case. Concretely, we found that the condition for brane-brane horizons to arise takes the form \eqref{conc_R_dot_tau}.
Hence, brane-brane horizons occur if the induced geometry on the boundary has cosmological horizons. 
It then follows that potentials satisfying
\begin{align}
    -\frac{\partial_\phi V}{V}<\lambda_{\rm crit}-\epsilon\,,\qquad \lambda_{\rm crit}=\frac{1}{\sqrt{d-2}}\,, \label{app_Horizon_critereon_bulk_field_general}
\end{align}
for $\phi>\phi_0,\epsilon>0$ induce brane-brane horizons whereas potentials satisfying
\begin{align}
    -\frac{\partial_\phi V}{V}\geq\lambda_{\rm crit}
    \label{sho}
\end{align}
do not. As before, one may focus on potentials for which the `measure of steepness' on the l.h.~side of \eqref{app_Horizon_critereon_bulk_field_general} and \eqref{sho} is constant. These are clearly exponential potentials, and the subset of those inducing brane-brane horizons is
\begin{align}
    V=V_0\exp(-\lambda \phi)\,
    \qquad\mbox{with}\qquad
    \lambda<\lambda_{\rm crit}\,.
    \label{conc_cond_bulk}
\end{align}
Allowing for asymptotically subleading terms, one finds that brane-brane horizons arise for potentials of the form
\begin{align}
    V=V_0\phi^\beta e^{-\lambda_{\rm crit}\phi}\,\qquad\mbox{with}\qquad \beta>\beta_{\rm crit}= \frac{1}{2}\,.\label{subl}
\end{align}
Intriguingly both the critical potentials of \eqref{conc_cond_bulk} and of \eqref{subl} correspond to the square roots of the analogous critical potentials in the standard FLRW setting without boundaries.

It would be interesting to study string-theoretic realizations of our ETW brane setups, the stringy origin of non-zero potentials for a bulk scalar and their dynamical effects, and the consequences of dimensional reduction for the Refined Horizon Criterion.

It remains a key open research question whether and precisely which cosmological horizons represent a fundamental inconsistency 
in quantum gravity. As part of this investigation, it would be important to find ETW branes in string theory satisfying \eqref{conclusion_brane_field} or \eqref{conc_cond_bulk}. As discussed above, it might be that since cosmological horizons (of brane-bulk type) exist in the Randall-Sundrum model, they are acceptable in general. This could be taken as support for a more optimistic view on de Sitter and inflation.

By contrast, if horizons are indeed {\it the} property responsible for the difficulties with de Sitter, a better understanding of our 
Refined Horizon Criterion is needed: Ideally, one would like to identify more fundamental features distinguishing the RS from a dS horizon.
One way forward may be analysing a `Causal-Isolation Criterion' that forbids cosmologies in which two inertial observers can be outside each other's cosmological horizon in their asymptotic future.
This criterion does not forbid the RS scenario: Here, the asymptotic future of the bulk observer lies outside the brane observer's horizon, while the bulk observer does not even possess a horizon. Another very interesting, even weaker version of a horizon-type criterion is the following:\footnote{
We thank the anonymous referee for pointing out this possibility.}

One could demand that, among the observers with infinite worldline, at least one should have no horizon. The conjecture would then be that all solutions of a consistent theory of quantum gravity must satisfy this condition. Again, the RS solution would not be ruled out since bulk observers have no horizon.

\subsection*{Acknowledgements} 
We would like to thank Thibaut Coudarchet and Jes\'us Huertas for helpful discussions. We are also grateful to Simon Schreyer and Victoria Venken for useful comments on the manuscript. This work was supported by Deutsche Forschungsgemeinschaft (DFG, German Research Foundation) under Germany’s Excellence Strategy EXC 2181/1 - 390900948 (the Heidelberg STRUCTURES Excellence Cluster).

\vspace{1cm}
\noindent
{\bf\Large Appendix}
\appendix
\section{Bulk Geometry}\label{App:A_B_u}
In this appendix, we study the bulk geometries of the spacetimes analyzed in Sect.~\ref{Sect:Brane_field} and Sect.~\ref{Sect:Bulk_field} in more detail.
In App.~\ref{App:A_B_u_brane} and App.~\ref{App:A_B_u_bulk}, we compute the profiles $A(u),B(u)$ (and $\phi(u)$ if present) for solutions that satisfy
\begin{align}
    R(\tau)=\tau^\alpha\,,
\end{align}
for the geometries studied in Sect.~\ref{Sect:flat_brane} and Sect.~\ref{Sect:Bulk_EoM} respectively.
For this purpose, we recall the relations \eqref{eigentime_Hor} between $t$ and $\tau$ and between $R$ and $B$:
\begin{align}
    d\tau = e^{A(t,z=0)}dt\,,\qquad R=e^{B(t,z=0)}\,.\label{app_Eigentime}
\end{align}
In App.~\ref{App:Embedding}, we study the embedding of the ETW brane of Sect.~\ref{Sect:flat_brane} in canonical Minkowski coordinates.
\subsection{Scalar Field living on the Brane}\label{App:A_B_u_brane}
We here consider $\phi$ to only live on the boundary, i.e.~we are considering the situation of Sect.~\ref{Sect:Brane_field}.
We can then compute at $z=0$
\begin{align}
		\partial_t B = e^A\partial_\tau B = e^A\frac{\alpha}{\tau} = e^{B/2}\sqrt{\partial_tB}\frac{\alpha}{\tau}\,,\label{app_Bp}
\end{align}
where we have used \eqref{sol_Einstein_brane}.
We thus find
\begin{align}
    \partial_t B = e^B\frac{\alpha^2}{\tau^2}=\alpha^2\tau^{\alpha-2}\,,
\end{align}
and by inserting this result back into \eqref{app_Bp}, one obtains
\begin{align}
    e^{A(\tau)}=\alpha \tau^{\alpha-1}\,.
\end{align}
Integrating \eqref{app_Eigentime} then yields the solution
\begin{align}
    t-u_0=\frac{1}{\alpha(2-\alpha)}\tau^{2-\alpha}\,\qquad \Leftrightarrow\qquad \tau=\left[\alpha(2-\alpha)(t-u_0)\right]^{\frac{1}{2-\alpha}}\,,
\end{align}
with $u_0$ an integration constant.
We can now easily determine the profiles for $A(t),B(t)$. Going away from $z=0$ simply results in replacing $t$ with $t-z=u$.
The corresponding profiles are:
\begin{align}
        e^A =\alpha \left[\alpha(2-\alpha)(u-u_0)\right]^{\frac{\alpha-1}{2-\alpha}}\,\,\,\,,\qquad e^B=\left[\alpha(2-\alpha)(u-u_0)\right]^{\frac{\alpha}{2-\alpha}}\,.\label{app_eA_eB}
\end{align}
\subsection{Embedding in canonical Minkowski Parametrization}\label{App:Embedding}
The metric \eqref{metric_Hor} with $A$ and $B$ related by \eqref{sol_Einstein_brane} is Riemann flat. 
In the following, we find the coordinate transformation that brings \eqref{metric_Hor} into standard Minkowski form and study how the ETW brane is embedded in this frame.

Using the relation \eqref{sol_Einstein_brane}, the metric \eqref{metric_Hor} takes the form
\begin{align}
    ds_g^2=-d(e^B)dv+e^{2B}ds_{d-2}^2 \equiv -d\tilde Rdv+\tilde R^2ds_{d-2}^2\,,
\end{align}
where we have defined $\tilde R(u)=e^{B(u)}$. Note that $\tilde R$ is equal to $R$ at $z=0$.
We denote the coordinates of $ds_{d-2}^2$ as $\Vec{x}$ and write $ds_{d-2}^2=d\vec x^2$.
We then define the new variables
\begin{align}
    \vec X = \tilde R\vec x\,,\qquad V=v+\frac{\vec X^2}{\tilde R}\,,\qquad T=\frac{V+\tilde R}{2}\,,\qquad w=\frac{V-\tilde R}{2}\,,
\end{align}
in which the metric becomes
\begin{align}
    ds_g^2=-d\tilde RdV+d\vec X^2=-dT^2+dw^2+d\vec X^2\,.
\end{align}
We clearly recognize flat space.
The ETW brane used to be at $u=v$, which translates to $\tilde R=e^{B(v)}$ and finally to 
\begin{align}
	T-w=\exp(B\left(T+w-\frac{\vec X^2}{T-w}\right))=\left[(2-\alpha)(T+w-\frac{\vec X^2}{T-w}-u_0)\right]^{\frac{\alpha}{2-\alpha}}\,,
\end{align}
where we have used \eqref{app_eA_eB}.
The above equation can be brought into the form (setting $u_0=0$ which can be done without loss of generality)
\begin{align}
    \vec X^2+w^2=T^2-\frac{(T-w)^\frac{2}{\alpha}}{2-\alpha}\,.
\end{align}
We see that the ETW brane, at late times where the last term above becomes negligible, is spherically embedded in the coordinates $\vec X,w$ and expands with the speed of light.
At finite time $T$, the embedding is not precisely spherical.
Nevertheless, this result sheds light on the question why we are finding the condition $\alpha>2$ also for the metric ansatz \eqref{metric_Hor}.
\subsection{Scalar Field living in the Bulk}\label{App:A_B_u_bulk}
In case the scalar field lives in the bulk, we can find the profiles of $A(u),B(u)$ as follows.
The nontrivial Einstein equation \eqref{Einstein_nontrivial} at $z=0$ can be written in terms of $\tau$ as
\begin{align}
    e^{2A}\left(\dot A\dot B -\frac{1}{2}(\dot{B}^2+\dot{A}\dot B +\ddot{B})\right)=e^{2A}\frac{\dot{\phi}^2}{2(d-2)}\,.\label{App_Einstein_bulk}
\end{align}
Using the fact that $\dot B=H$ and \eqref{bdry_eqs_Hor}, one finds \begin{align}
    \ddot B = \dot H = \dot\phi \frac{\partial_\phi V}{M_d^{d-2}(d-2)}=-\frac{\dot\phi^2}{(d-2)}\,,
\end{align}
such that \eqref{App_Einstein_bulk} reduces to
\begin{align}
    \dot{A}=\dot{B}\,,\qquad\Rightarrow\qquad A=B\,.\label{app_A_equal_B}
\end{align}
The integration constant can be set to zero by a redefinition of $v$.
For profiles of the form $R=\tau^\alpha$ it directly follows that
\begin{align}
    A=B=\alpha\log{\tau}\,.
\end{align}
Integration of \eqref{app_Eigentime} then results in 
\begin{align}
    t-u_0=-\frac{1}{\alpha-1}\tau^{-(\alpha-1)}\,,\qquad \Rightarrow\qquad \tau=\left[-(\alpha-1)(t-u_0)\right]^{-\frac{1}{\alpha-1}}\,.
\end{align}
One thus finds the profiles
\begin{align}
    e^{A(u)}&=e^{B(u)}=\left[-(\alpha-1)(u-u_0)\right]^{-\frac{\alpha}{\alpha-1}}\,,\\
    \phi(u)&=-\frac{\sqrt{(d-2)\alpha}}{\alpha-1}\log(-\frac{V_0}{(d-2)\alpha}\left[-(\alpha-1)(u-u_0)\right]^{-\frac{1}{\alpha-1}})\,.
\end{align}

\section{Recovering the Potential from the Expansion Rate}\label{App:H_R_V_phi}
In this section, we explain how to compute $V(\phi)$ if $H(\tau)$ or $H(R)$ is given.
A similar analysis has been performed in \cite{Starobinsky:1998fr,Copeland:2006wr}.
We restrict our attention to expanding universes, i.e.~$H>0$, and to scalar field profiles that everywhere satisfy $\dot\phi > 0$.

If $H(\tau)$ is known, one can find the profile $H(R)$ by first obtaining $R(\tau)$ from solving the differential equation
\begin{align}
    \dot R(\tau) = R(\tau)\cdot H(\tau)\,,
\end{align}
and then inverting this function to get $\tau(R)$, which then straightforwardly gives $H(\tau(R))$.
Due to the assumption $H>0$, the above procedure is well defined.
\subsection{Scalar Field living on the Brane}\label{App:H_R_V_phi_brane}
In this section, we consider the system of equations given by the Friedmann equation of the form \eqref{Friedmann_brane_field} or \eqref{Friedmann_brane_flat} and the scalar field equation \eqref{brane_eom}.

The Friedmann equation then straightforwardly determines the profile $\rho(R)$ of the energy density.
Using the continuity equation \eqref{continuity}, it is easily shown that $\rho$ satisfies the relation
\begin{equation}
    \frac{d \rho}{dR} = \frac{1}{HR} \frac{d \rho}{d \tau} =-2(d-2)M_d^{d-3}\frac{T}{R},
\end{equation}
where $T=\dot\phi^2/2$ is the kinetic energy of the scalar field. From this relation, the profile $T(R)$ of the kinetic energy directly follows.
The scalar field profile $\phi(R)$ follows from solving
\begin{align}
    \frac{d \phi}{d R} = \frac{1}{HR} \frac{d \phi}{d \tau} = \frac{\sqrt{2T(R)}}{HR}\,.
\end{align}
The corresponding function $\phi(R)$ can be inverted to obtain $R(\phi)$, from which the potential simply follows as
\begin{equation}
    V(\phi) = \rho(R(\phi)) - T(R(\phi)).
\end{equation}
Thus, given any strictly positive expansion rate $H(\tau)$ or $H(R)$, the corresponding potential $V(\phi)$ realizing it can be computed. 

\subsection{Scalar Field living in the Bulk}\label{App:H_R_V_phi_bulk}
In case the defining system of equations is given by \eqref{bdry_eqs_Hor}, the procedure of obtaining $V(\phi)$ is as follows.
The Friedmann equation directly determines $V(R)$.
From the second equation in \eqref{bdry_eqs_Hor}, it follows that
\begin{align}
    \left(\frac{d\phi}{dR}\right)^2=-HR\frac{dV}{dR}\,,
\end{align}
from which $\phi(R)$, and by inversion $R(\phi)$, can be obtained. 
The potential $V(\phi)$ is then simply given by
\begin{align}
    V(\phi)=V(R(\phi))\,.
\end{align}

\section{Autonomous System Analysis}\label{App:Critical_cases}
In this appendix, we set $M_d=1$.
For a review on the usage of autonomous systems in cosmology, see \cite{Bahamonde:2017ize} and references therein.
\subsection{Scalar Field living on the Brane}\label{App:Critical_brane}
In Section \ref{Sect:Brane_field}, we showed that brane-brane horizons exist for potentials
of the form
\begin{align}
    V(\phi)=\frac{c^2}{\phi^2}\,,\qquad \text{with }\qquad c>c_{\rm crit}\equiv \sqrt{2(4d-9)}\,,\label{app_brane_critical_V}
\end{align}
and that they do not exist if $c\leq c_{\rm crit}$.
For \eqref{app_brane_critical_V} with $c=c_{\rm crit}$, no brane-brane horizons can occur, but it is clear from \eqref{R_alpha_g_2_flat} that they are only marginally disallowed as $R$ is given by $R\sim\tau^2$.
We now study the conditions for brane-brane horizons to occur in more detail, including the case when $V(\phi)$ asymptotes to the critical potential
\begin{align}
    V_{\rm crit}=\frac{c_{\rm crit}^2}{\phi^2} \label{V_crit}
\end{align}
for $\phi\to\infty$.

For the flat embedding considered in Sect.~\ref{Sect:flat_brane}, the condition for brane-brane horizons to arise is that the coordinate $t$ at the ETW brane is bounded from above. 
This condition can be written as
\begin{align}
    \infty > \int^{t_{\rm max}} dt = \int^{\infty} e^{-A(t)}d\tau=\int d\tau\frac{1}{e^{-A(t)}e^{B(t)}\partial_t B(t)}=\int^\infty\frac{d\tau}{\dot R(\tau)}\,,\label{app_cond_hor}
\end{align}
where $t_{\rm max}$ is the maximal value of $t$ at the ETW brane (including the possibility of $t_{\rm max}=\infty$) and where we used \eqref{sol_Einstein_brane} and \eqref{eigentime_Hor}.
This is the same condition as \eqref{brane_brane_condition_Rdot}, which we found for the spherical brane embedding.
We now proceed with studying the consequences of this condition on scalar potentials.

To study the system of equations \eqref{Friedmann_brane_flat},\eqref{brane_eom} in greater detail, we first introduce a new time variable $\tilde\tau$ through
\begin{align}
    d\tilde\tau = Hd\tau\,,\label{tau_tilde_def}
\end{align}
leading to
\begin{align}
    \frac{dR}{d\tilde{\tau}}=\frac{\dot R}{H}=R\,,\qquad\Rightarrow\qquad R=\text{const}\cdot\exp(\tilde{\tau})\,.
\end{align}
The condition \eqref{app_cond_hor} then takes the form\footnote{If $H(\tau)$ behaves to leading order as $H\sim\alpha/\tau$, then $\tilde{\tau}(\tau\to\infty)=\infty$.} 
\begin{align}
    \infty >\int^\infty \frac{d\tau }{\dot R(\tau)}=\int^\infty \frac{d\tau }{HR}=\int^\infty \frac{d\tilde{\tau}}{H^2R}=\text{const}\cdot\int^\infty d\tilde{\tau} \frac{e^{-\tilde{\tau}}}{\rho^2}\,,\label{horizion_condition_appendix}
\end{align}
where we have used the Friedmann equation \eqref{Friedmann_brane_flat} in the final step. 
We will study the $\tilde{\tau}$-dependence of $\rho$ in the following. 
We introduce the dynamical parameter
\begin{align}
    \delta(\phi)=\partial_\phi \frac{1}{\sqrt{V(\phi)}}\,,
\end{align}
which can be seen as a measure for the local steepness of the potential.
We note that for potentials of the form \eqref{app_brane_critical_V}, $\delta = 1/c$. 
By considering the variables $\delta$ and $U\equiv \dot\phi/\sqrt{2\rho}$, the dynamics of the system defined by \eqref{Friedmann_brane_flat} and \eqref{brane_eom} (together with \eqref{tau_tilde_def}) can be studied via an autonomous system of equations taking the form (see e.g.~\cite{Mizuno:2002wa,Mizuno:2004xj,Goliath:1998na,Zhou:2007xp})
\begin{align}
    \frac{d U}{d \tilde{\tau}} &= (d-2) (1-U^2) \left( \sqrt{2} \sqrt{1 - U^2} \delta - U\right)\,,\label{U_tildetau}\\
    \frac{d \delta}{d \tilde{\tau}} &= \sqrt{2} (d-2) f(\delta) U \sqrt{1 - U^2}, \qquad f(\delta) \equiv \frac{\partial_\phi \delta}{\sqrt{V}}\,.
\end{align}
If $f(\delta)$ vanishes at some $\delta=\delta_*$, there exists a fixed point given by\footnote{In \cite{Escobar:2013js,Fadragas:2013ina}, the function $f(\delta)$ was named $f$-deviser.}
\begin{align}
    \delta=\delta_*\,,\qquad   U=U_*=\sqrt{\frac{2}{2\delta_*^2+1}}\delta_*\,.\label{FP_brane_general}
\end{align}
For potentials of the form \eqref{app_brane_critical_V}, one simply finds $\delta_*=1/c$, as $f(\delta)$ vanishes identically.

The profile $\rho(\tilde{\tau})$ of the energy density, needed to evaluate the integral in \eqref{horizion_condition_appendix}, can be determined from the continuity equation \eqref{continuity}, which takes the form
\begin{equation}
    \frac{1}{\rho} \frac{d \rho}{d \tilde \tau} = -2 (d-2) U^2.\label{app_rho_evol}
\end{equation}
It is then straightforward to see that the condition \eqref{horizion_condition_appendix} is fulfilled if for large $\tilde\tau$, $U$ satisfies
\begin{align}
    U<U_{\rm crit}-\epsilon \equiv \frac{1}{2\sqrt{d-2}}-\epsilon\,,\qquad \text{for some}\qquad \epsilon>0\,.\label{App_U_crit}
\end{align}
If the fixed point \eqref{FP_brane_general} obeys $U_*<U_{\rm crit}$, which by using \eqref{FP_brane_general} is equivalent to $\delta<1/c_{\rm crit}$, it follows from \eqref{App_U_crit} then brane-brane horizons will exist. 
Furthermore, if $\delta<1/c_{\rm crit}-\epsilon$ for $\phi>\phi_0$, with arbitrary constants $\epsilon$ and $\phi_0$, Eq.~\eqref{U_tildetau} directly implies that $U$ will stay strictly below $U_{\rm crit}$ for $\tilde\tau\to\infty$, such that brane-brane horizons will occur.
Similarly, if $\delta\geq 1/c_{\rm crit}$ for $\phi>\phi_0$, it follows that brane-brane horizons can not arise, as $U$ will asymptotically be too large.\footnote{For $\delta=1/c_{\rm crit}$, Eq.~\eqref{U_tildetau} states that if $U$ is initially small, it grows and approaches $U_{\rm crit}$.
We know it does so fast enough that no horizons will form. If now $\delta\geq 1/c_{\rm crit}$ and $U$ is initially small, Eq.~\eqref{U_tildetau} implies that $U$ grows at least as fast as before, such that again no horizons can form.}
This shows that the conditions \eqref{Class_brane_field} and \eqref{cvc} can indeed determine the presence or absence of brane-brane horizons.

We now turn to the study of potentials that are critical to leading order but exhibit a subleading correction making the total potential slightly flatter.
In other words, we consider potentials for which $\delta(\phi)$ approaches $1/c_{\rm crit}$ from below for $\phi\to\infty$ such that the fixed point \eqref{FP_brane_general} takes the form
\begin{align}
U_* = \frac{1}{2\sqrt{d-2}}\,,\qquad \delta_* = \frac{1}{c_{\rm crit}}=\frac{1}{\sqrt{2(4d-9)}}\,.\label{FP_sub}
\end{align}
We are interested in conditions on the corrections to $V_{\rm crit}$ imposed by \eqref{horizion_condition_appendix}.
The form of $f(\delta)$ determines how quickly the potential approaches the critical one.
We focus on potentials for which $f(\delta)$ is analytic and quadratic at leading order
\begin{align}
    f(\delta) \simeq \frac{1}{2 \eta} \left(\frac{1}{c_{\rm crit}} - \delta\right)^2\,.
\end{align}
This form of $f(\delta)$ is (to leading order) obtained from potentials of the form\footnote{By contrast, functions that vanish to linear order $f(\delta)\sim (1/c_{\rm crit}-\delta)$ are obtained from potentials of the form $V=c_{\rm crit}^2/\phi^2(1+\eta/\phi^n)$ for $n>0$ and can never lead to the formation of brane-brane horizons.}
\begin{align}
    V(\phi)=\frac{c_{\rm crit}^2}{\phi^2}\left(1+\frac{\eta}{\ln(\phi)}\right)\,.\label{V_crit_subleading}
\end{align}
Clearly, for $\phi\to\infty$, \eqref{V_crit_subleading} approaches \eqref{V_crit}.
As a result of considering a quadratic $f(\delta)$, the fixed-point \eqref{FP_sub} is non-hyperbolic.
We consider the new variables
\begin{align}
    u = U-U_*\,,\qquad\kappa = \delta-\delta_*\,,
\end{align}
and then introduce
\begin{align}
    y = u - \frac{c_{\rm crit}^3}{16 (d-2)^{3/2}} \kappa\,.
\end{align}
In the variables $y,\kappa$, the equations of motion are diagonal to leading order
\begin{align}
    \frac{d y}{d\tilde\tau} &= -(d-2) y + h(\kappa) + \mathcal{O}(\kappa^3, y^2, \kappa y), \quad h(\kappa) = \mathcal{O}(\kappa^2)\,,\\
    \frac{d \kappa}{d \tilde \tau} &= \frac{c_{\rm crit}}{8 \eta} \kappa^2 + \mathcal{O}(\kappa^3, \kappa^2 y)\,.\label{kappa_eom}
\end{align}
A general result for autonomous systems, one of the center manifold theorems, see e.g.~\cite{perko2013differential}, implies that the solution for $y$ satisfies $y = \mathcal{O}(\kappa^2)$. Therefore, it is safe to neglect the terms $\order{\kappa^2y}$ in \eqref{kappa_eom} and the asymptotic solutions for $\kappa$ and $u$ can be determined:
\begin{align}
    y&=\order{\tilde\tau^{-2}}\,,\qquad \kappa = -\frac{8 \eta}{c_{\rm crit} \tilde\tau} + \mathcal{O}(\tilde{\tau}^{-2})\,,\label{kappa_tau_tilde}\\
    u &= y + \frac{c_{\rm crit}^3}{16 (d-2)^{3/2}} \kappa = -\frac{\eta c_{\rm crit}^2}{ 2 (d-2)^{3/2} \tilde\tau} + \mathcal{O}(\tilde{\tau}^{-2})\,.\label{U_tau_tilde}
\end{align}
The asymptotic form of the energy density $\rho$ follows from inserting the asymptotic field profile \eqref{U_tau_tilde} into \eqref{app_rho_evol}
\begin{align}
    \frac{1}{\rho}\frac{d \rho}{d \tilde\tau} = - \left(\frac{1}{2} - \frac{c_{\rm crit}^2 \eta}{(d-2) \tilde\tau} + \mathcal{O}(\tilde{\tau}^{-2}) \right)\,,\qquad \Rightarrow\qquad \rho = \text{const} \cdot \tilde{\tau}^{\frac{\eta c_{crit}^2}{d-2}} e^{-\tilde\tau + \mathcal{O}(\tilde{\tau}^{-1})}\,.\label{V_tau_tilde}
\end{align}
With this asymptotic solution, the condition \eqref{horizion_condition_appendix} takes the form
\begin{align}
    \infty > \int^\infty d\tilde{\tau} \tilde{\tau}^{-\frac{\eta c_{crit}^2}{d-2}} \,,
\end{align}
from which we conclude that brane-brane horizons exist if
\begin{align}
    \eta > \eta_{\rm crit} \equiv \frac{d-2}{c_{crit}^2}=\frac{d-2}{2(4d-9)} \,,
\end{align}
and that they do not if $\eta\leq\eta_{\rm crit}$.
For potentials \eqref{V_crit_subleading} with $\eta=\eta_{\rm crit}$, one could perform a similar analysis, by looking at functions $f(\delta)$ that vanish to higher than quadratic order, to determine which subleading corrections to it lead to the existence of brane-brane horizons. 

\subsection{Scalar Field living in the Bulk}\label{App:Critical_bulk}
We now consider the situation of the scalar field $\phi$ living in the bulk.
The condition for horizons to exist is then given by
\begin{align}
    \infty > \int^{t_{\rm max}} dt &= \int^\infty e^{-A}d\tau = \int^\infty e^{-B}d\tau = \int^\infty \frac{d\tau}{R}= \int^\infty \frac{e^{-\tilde{\tau}}}{H}d\tilde{\tau}=\text{const.}\int^\infty \frac{e^{-\tilde{\tau}}}{V}d\tilde{\tau}\label{app_condition_R_bulk}\,,
\end{align}
where we have used \eqref{app_A_equal_B}, the definition \eqref{tau_tilde_def} and the Friedman equation \eqref{bdry_eqs_Hor}.
Similar to the case where the field solely lives on the brane, there exists an autonomous system description for the equations \eqref{bdry_eqs_Hor}.
As the scalar field equation here is first order, the autonomous system is formed of one less equation than before.
We define the dynamical parameter 
\begin{align}
    \zeta = -\frac{\partial_\phi V}{V}\,,
\end{align}
which is simply equal to $\lambda$ in case the potential takes the form \eqref{potential_bulk_field}.
The evolution equation in terms of the variable $\tilde\tau$, defined in \eqref{tau_tilde_def}, is given by
\begin{align}
    \frac{d\zeta}{d\tilde{\tau}}= (d-2) f(\zeta)\,,\qquad f(\zeta)\equiv \zeta\frac{d\zeta}{d\phi}\,.\label{zeta_tautilde}
\end{align}
A fixed point $\zeta_*$ arises at zeros of the $f$
\begin{align}
    f(\zeta_*)=0\,.
\end{align}
For potentials of the form \eqref{potential_bulk_field}, a fixed point is given by $\zeta_*=\lambda$.
Using \eqref{bdry_eqs_Hor}, one finds the evolution equation of $V$ (and hence $H$) to be
\begin{equation}
    \frac{1}{V} \frac{d V}{d \tilde\tau} = -(d-2) \zeta^2 \,.\label{App_V_tau_bulk}
\end{equation}
By integrating the above and plugging the result into \eqref{app_condition_R_bulk}, it follows that \eqref{app_condition_R_bulk} is fulfilled in case $\zeta<\lambda_{\rm crit}-\epsilon$ holds for $\phi>\phi_0$, where $\epsilon>0$ and $\phi_0$ are arbitrary constants.
Similarly, if $\zeta \geq\lambda_{\rm crit}$ holds for $\phi>\phi_0$, then brane-brane horizons can not exist.
This shows that the conditions \eqref{Horizon_critereon_bulk_field_general} and \eqref{Horizon_critereon_bulk_field_general_no_horizon} can determine the presence of absence of brane-brane horizons.

We are now interested in situations where the potential is to leading order given by the exponential 
\begin{align}
    V=V_0\exp(-\lambda_{\rm crit}\phi)\,,\qquad \lambda_{\rm crit}=\frac{1}{\sqrt{d-2}}\,,
\end{align}
or, in other words, in functions $f(\zeta)$ that vanish for
\begin{align}
    \zeta_*=\lambda_{\rm crit}=\frac{1}{\sqrt{d-2}}\,.
\end{align}
As before, we are interested in situations where $f(\zeta)$ is (to leading order) quadratic
\begin{align}
    f(\zeta)=\frac{\lambda_{\text{crit}}}{\beta}(\zeta-\zeta_*)^2\,,
\end{align}
which occurs for potentials of the form
\begin{align}
    V=V_0\phi^\beta e^{-\lambda_{\rm crit}\phi}\,.
\end{align}
The leading order solution to \eqref{zeta_tautilde} is then given by
\begin{align}
    \zeta=\zeta_*\left(1 - \frac{\beta}{\tilde\tau}\right)\,.
\end{align}
Using \eqref{App_V_tau_bulk}, one finds
\begin{align}
    V\sim e^{-\tilde{\tau}}\tau^{2\beta}\,,
\end{align}
which, inserted into \eqref{app_condition_R_bulk}, yields
\begin{align}
    \infty > \int^\infty d\tilde{\tau}\; \tilde{\tau}^{-2\beta}\,.
\end{align}
One directly reads off the condition for the existence of brane-brane horizons to be
\begin{align}
    \beta > \frac{1}{2}\,.
\end{align}
If $\beta \leq 1/2$, brane-brane horizons do not occur.

\subsection{FLRW Universes}\label{App:Critical_closed}
In this section we consider the dynamics of an FLRW universe without boundary, following from the action
\begin{align}
    S=\frac{1}{2}\int_\mathcal{M}d^dx\sqrt{-g}\left(\mathcal{R}_d-g^{\mu\nu}\partial_\mu\phi\partial_\nu\phi-2V(\phi)\right)\,,
\end{align}
with metric of the form
\begin{align}
    ds^2=-d\tau^2+a(\tau)^2ds_{d-1}^2\,.
\end{align}
We here consider the spatial metric $ds_{d-1}^2$ to be flat, e.g.~the one of a torus.\footnote{We expect the same critical potentials to arise when considering spatial geometries with nonzero curvature.} 
We consider $V(\phi)$ to be monotonically decaying to zero from above for $\phi\to\infty$.
The equations of motion are the ordinary Friedmann equations and the scalar field equation for cosmology
\begin{align}
    H^2&=\left(\frac{\dot a}{a}\right)^2=\frac{2\rho}{(d-1)(d-2)}\,,\qquad \rho = \frac{\dot\phi^2}{2}+V(\phi)\,,\qquad \ddot\phi+(d-1)H\dot\phi = -\partial_\phi V\,.
\end{align}
It is well known that potentials of the form 
\begin{align}
    V=V_0\exp(-\gamma\phi)\,,\qquad \text{with}\qquad \gamma<\gamma_{\rm crit}\equiv\frac{2}{\sqrt{d-2}}
\end{align}
realize cosmological horizons.
Using the definition \eqref{tau_tilde_def} and the Friedmann equation, the condition \eqref{horizon_calc} takes the form
\begin{equation}
    \infty > \int^\infty \frac{d \tau}{a(\tau)} = \int^\infty \frac{e^{-\tilde\tau} d \tilde\tau}{H} = \text{const.} \int^\infty \frac{e^{-\tilde\tau} d \tilde\tau}{\sqrt{\rho}}\,.\label{app_one_over_R_closed}
\end{equation}
To study situations where the potential is to leading order of the form $V=V_0\exp(-\gamma_{\rm crit}\phi)$, we consider the dynmaical variables
\begin{align}
    U = \frac{\dot{\phi}}{\sqrt{2 \rho}}\,,\qquad \text{and}\qquad \xi = - \frac{\partial_\phi V}{V}\,.
\end{align}
Using the definition \eqref{tau_tilde_def}, we find the autonomous system of equations (see e.g.~\cite{Bahamonde:2017ize,Copeland:1997et})
\begin{align}
    \frac{dU}{d \tilde\tau} &= (d-1)\left(1 - U^2\right) \left( \frac{1}{2} \sqrt{\frac{d-2}{d-1}} \xi - U\right)\,,\label{U_tildetau_closed}\\
    \frac{d\xi}{d \tilde\tau} &= \sqrt{(d-1)(d-2)} f(\xi) U\,, \qquad f(\xi) \equiv \frac{d\xi}{d\phi}\,.
\end{align}
The energy density satisfies the relation (following from the continuity equation)
\begin{align}
    \frac{1}{\rho} \frac{d \rho}{d\tilde\tau} = -2(d-1) U^2\,,\label{app_closed_rho}
\end{align}
from which, together with \eqref{app_one_over_R_closed}, it follows that cosmological horizons exist as long as $U$ asymptotically stays strictly below $1/\sqrt{d-1}$.
Equation \eqref{U_tildetau_closed} shows that this is guaranteed to be the case if $\xi$ stays strictly below $\gamma_{\rm crit}$.
Similarly, if $\xi\geq \gamma_{\rm crit}$ for $\phi>\phi_0$, cosmological horizons will not exist.

To study potentials that are critical to leading order, we again consider functions $f(\xi)$ that vanish quadratically at $\xi=\gamma_{\rm crit}$
\begin{equation}
    f(\xi) = \frac{1}{\beta} \left( \xi - \gamma_{\text{crit}}\right)^2\,.
\end{equation}
Such a functional form is obtained from potentials of the form 
\begin{equation}
    V = V_0 \phi^{\beta} e^{-\gamma_{\text{crit}}\phi}\,.
\end{equation}
As before, we study the solutions near the non-hyperbolic fixed point, here given by
\begin{equation}
     U_* \equiv \frac{1}{2} \sqrt{\frac{d-2}{d-1}} \gamma_{\text{crit}}=\frac{1}{\sqrt{d-1}}\,, \qquad\, \xi_* \equiv \gamma_{\text{crit}}=\frac{2}{\sqrt{d-2}}\,.
\end{equation}
Introducing the coordinates
\begin{equation}
    u \equiv U - U_*\, , \qquad \kappa \equiv \xi - \xi_* \qquad \text{and} \qquad y \equiv u - \frac{1}{2} \sqrt{\frac{d-2}{d-1}} \kappa \, ,
\end{equation}
the leading order equations of motion take the form
\begin{align}
    \frac{d y}{d \tilde\tau} &= -(d-2) y + h(\kappa) + \mathcal{O}\left( \kappa^3, y^2, \kappa y\right) \,,\qquad h(\kappa)=\order{\kappa^2}\,,\\
    \frac{d \kappa}{d\tilde\tau} &= \frac{\sqrt{d-2} \kappa^2}{\beta} + \mathcal{O}(\kappa^3, \kappa^2 y)\,.
\end{align}
From the center-manifold theorems (see e.g.~\cite{perko2013differential}), it follows that $y=\order{\kappa^2}$ such that the leading order solutions can be determined to be
\begin{equation}
    \kappa = - \frac{\beta}{\sqrt{d-2} \tilde\tau} + \order{\tilde\tau^{-2}}\,,  \qquad u = - \frac{\beta}{2 \sqrt{d-1} \tilde\tau}+ \order{\tilde\tau^{-2}}\,.
\end{equation}
Finally, we can compute the profile of the energy density from \eqref{app_closed_rho}
\begin{equation}
    \rho = \text{const.} \cdot \tilde{\tau}^{2 \beta} e^{-2\tilde\tau}.
\end{equation}
Thus, by plugging the result into \eqref{app_one_over_R_closed}, it follows that cosmological horizons exist if
\begin{equation}
    \beta > 1 \,,
\end{equation}
and that they do not if $\beta\leq 1$.

\bibliographystyle{utphys}
\bibliography{refs}

\end{document}